%% file: 2026-ieeevr-old-new-animation-system.tex
\documentclass{vgtc}                          

\graphicspath{{figures/}{pictures/}{images/}{./}} 

\usepackage{times}                     

\usepackage{tabu}                      
\usepackage{booktabs}                  
\usepackage{lipsum}                    
\usepackage{mwe}                       
\usepackage{multirow}                   
\usepackage{upgreek}
\usepackage{graphicx}
\usepackage{subcaption}
\usepackage{adjustbox}
\usepackage{xspace}
\AtBeginDocument{%
}

\usepackage{mathptmx}                  

\newcommand{\unitdegree}[1]{{#1}°}

\newcommand{\unitmillisecond}[1]{{#1}\,ms}

\newcommand{\unitminute}[1]{{#1}\,min}

\newcommand{\unithertz}[1]{{#1}\,Hz}
\newcommand{\unitframespersecond}[1]{{#1}\,fps}

\newcommand{\unitgram}[1]{{#1}\,g}

\newcommand{\enquote}[1]{``{#1}''}

\DeclareRobustCommand{\ivr}{IVR\xspace}

\onlineid{1474}

\vgtccategory{Research}

\vgtcinsertpkg

\title{Technological Advances in Two Generations of Consumer-Grade VR Systems: Effects on User Experience and Task Performance}

\author{%
\begin{tabular}{ccc}
\parbox{1.4in}{\centering
Marie Luisa Fiedler\thanks{These authors have contributed equally to this work} $^{ \dag}$\\
\scriptsize HCI \& PIIS Group\\ University of Würzburg
} &
\parbox{1.4in}{\centering
Christian Merz$^{*}$\thanks{e-mail: firstname.lastname@uni-wuerzburg.de}\\
\scriptsize HCI \& PIIS Group\\ University of Würzburg
} &
\parbox{1.4in}{\centering
Jonathan Tschanter$^{* \dag}$\\
\scriptsize HCI \& PIIS Group\\ University of Würzburg
}
\\[1.2em] 
\multicolumn{3}{c}{%
\begin{tabular}{cc}
\parbox{1.4in}{\centering
Carolin Wienrich$^{\dag}$\\
\scriptsize PIIS Group\\ University of Würzburg
} &
\parbox{1.4in}{\centering
Marc Erich Latoschik$^{\dag}$\\
\scriptsize HCI Group\\ University of Würzburg
}
\end{tabular}}
\end{tabular}%
}

\teaser{
  \centering
  \includegraphics[width=\linewidth]{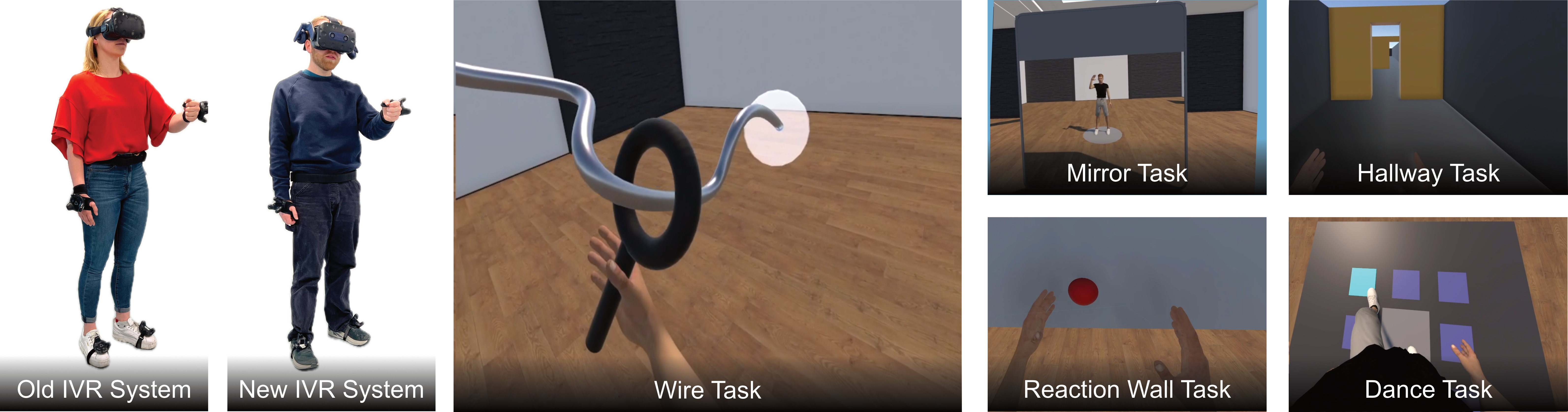}
  \caption{
  On the left, the figure shows the two generations of integrated virtual reality systems used to reconstruct users' movements and animate an avatar in the virtual environment.
  On the right, it presents the five tasks participants completed in our user-centered evaluation.
  }
  \label{fig:teaser}
}

\abstract{
    \input{sections/00-abstract}
} 

\keywords{Avatar, virtual reality, sense of embodiment, presence, immersion, task load.}

\usepackage{enumitem}
\setlist[itemize]{label={}, topsep=0pt} 

\begin{document}



\maketitle
\input{sections/01-introduction}
\input{sections/02-related-work}
\input{sections/03-method}
\input{sections/04-result}
\input{sections/05-discussion}
\input{sections/06-conclusion}

\acknowledgments{We thank Joanna Grause for her help in the study conduct, and Veronika Wirt for her help in the study implementation and study conduct. This research has been funded by the Bavarian State Ministry for Digital Affairs in the project XR Hub (A5-3822-2-16).}

\bibliographystyle{abbrv-doi-narrow}

\bibliography{references}


\end{document}

%% file: sections/00-abstract.tex
Integrated VR (IVR) systems consist of a head-mounted display (HMD) and body-tracking capabilities.
They enable users to translate their physical movements into corresponding avatar movements in real-time, allowing them to perceive their avatars via the displays.
Consumer-grade IVR systems have been available for 10 years, significantly fostering VR research worldwide. However, the effects of even apparently significant technological advances of IVR systems on user experience and the overall validity of prior embodiment research using such systems often remain unclear. 
We ran a user-centered study comparing two comparable IVR generations: a nearly 10-year-old hardware (HTC Vive, 6-point tracking) and a modern counterpart (HTC Vive Pro 2, 6-point tracking).
To ensure ecological validity, we evaluated the systems in their commercially available, as-is configurations. 
In a 2x5 mixed design, participants completed five tasks covering different use cases on either the old or new system. 
We assessed presence, sense of embodiment, appearance and behavior plausibility, workload, task performance, and gathered qualitative feedback.
Results showed no significant system differences, with only small effect sizes. Bayesian analysis further supported the null hypothesis, suggesting that the investigated generational hardware improvements offer limited benefits for user experience and task performance. For the 10-year generational step examined here, excluding potential technological progress in the necessary software components, this supports the validity of conclusions from prior work and underscores the applicability of older configurations for research in embodied VR.

%% file: sections/01-introduction.tex
\section{Introduction}
\label{sec:introduction}

Integrated VR (\ivr) systems enable highly immersive and embodied interactive experiences in real-time.
\ivr systems are characterized by a combination of a Head-Mounted Display (HMD) providing both display and head pose tracking, and additional sensor capabilities for tracking controllers, specific user limbs, or arbitrary objects at or around the users.
Since their widespread availability as consumer-grade products in the last 10 years, \ivr systems are used in many application areas, from games to, e.g., face-to-face-like interactions~\cite{smith2018communicationbehaviourvr, Young2015Dyadicinteractionsavatars}, training simulations~\cite{caserman2022impactfullbodytrainingpolice}, to therapy contexts~\cite{Bartl2022EffectsAvatarEnvironment, Hepke2024DevelopmentValidation3D,portingale2024review}.
Overall, in combination with the arrival of powerful yet easily accessible software systems supporting XR developments (XR short for eXtended Reality; encompassing Augmented, Mixed, and Virtual Reality) like Unity 3D or the Unreal Engine, consumer-grade \ivr systems fostered a significant surge in XR applications and research due to availability and low costs.

The quality of \ivr systems can be determined by their degree of immersion, defined by Slater~\cite{slater1997immersionandpresence}:
\enquote{The more that a system delivers displays (in all sensory modalities) and tracking that preserves fidelity to their equivalent real-world sensory modalities, the more that it is 'immersive'} \cite{slater2003note}.
Immersion, in turn, affects the user's perception of the virtual environment and the overall experience~\cite{cummings2016immersive}. 
For example, prior research has shown the influence of system characteristics on user experience~\cite{cummings2016immersive, juan2009comparison}, sense of embodiment towards the avatar~\cite{born2019exergamesimmersionembodiment, eubanks2020bodytrackingfidelityik, goncalves2022impactbodytrackingsenseofembodiment, yun2023animationfidelityselfavatar}, or task performance~\cite{yun2023animationfidelityselfavatar}.
However, technical characteristics of \ivr systems are particularly system-specific.
They potentially differ in many aspects, from low-level tracking technologies (e.g., optical markers, inertial sensors, depth cameras), to the application and concrete implementation of procedural or physics-based models and animation algorithms (e.g., inverse kinematics, motion capture retargeting, or data-driven synthesis), also resulting in quite different system specs, e.g., in refresh rate, spatial resolution, field of view, latency, and tracking fidelity.
Such differences in technical characteristics and, hence, immersive capabilities are often notably pronounced between different generations of systems due to the overall technological progress.
However, many empirical findings on XR-specific user experience, e.g., on presence or avatar embodiment, are based on the factual immersive characteristics of one specific system.
Taking into account the impact of technical characteristics of \ivr systems on user experience in combination with the potentially distinct differences of such characteristics, specifically between different technological generations, raises questions of reproducibility and generalizability.
We treat replication of empirical findings on multiple \ivr systems as a step toward ecological validity: 
ensuring that (also prior) user experience findings generalize beyond a single system to \ivr devices with substantially different technology characteristics in real-world use.
Thereby, we motivate a user-centered, rather than a hardware-centered approach, specifically across different system generations with largely differing technological characteristics.
This raises the following research question:

\begin{itemize}
    \item \textbf{RQ:} How do different generations of \ivr systems influence the user experience?
\end{itemize}

We address this gap by comparing user experience across two generations of \ivr systems: a nearly 10-year-old setup (HTC Vive with 6-point-tracking) and a state-of-the-art counterpart (HTC Vive Pro 2 with 6-point-tracking).
We adopted a holistic approach, evaluating each system in its commercially available as-is configuration. This choice prioritizes ecological validity by reflecting how most end-users and researchers encounter these systems. However, we used the exact same modern software stack for the kinematic solver and the overall simulation system due to well-known replication obstacles~\cite{grubel:2023a}. 
Our comparison, therefore, spans two generations of \ivr systems, where both the full-body animation system (via different tracker configurations) and the display (via different HMDs) may influence user experience.
We conducted a user-centered study using a $2\times5$ mixed-design, with participants performing five VR tasks designed to simulate different use cases and probe different aspects of \ivr systems' performance.
Results showed no significant differences between the systems, with Bayesian analyses supporting the null hypothesis of negligible benefits from generational improvements.
The findings suggest that, despite notable technological advancements, newer \ivr systems provide little perceptible advantage in user experience and task performance.
For the 10-year generational step examined here and excluding any potential technological progress in the necessary software parts, this supports the validity of conclusions from prior work and underscores the applicability of older configurations for research in embodied VR.

%% file: sections/02-related-work.tex
\section{Related Work}
Full-body animations have been researched for more than three decades \cite{Badler1993RealTimeControl, Boulic1997Integrationmotioncontrol}.
They have evolved from initial efforts to track a full-body pose into highly available consumer-grade \ivr systems for real-time avatar embodiment.
An \ivr system (1) reconstructs the movements of a user, typically by combining sensors, trackers, cameras, and software, without necessarily tracking every single body part (e.g., by using sparse tracking solutions); (2) maps them on a virtual avatar, and (3) displays the embodied avatar with an HMD. 
Different characteristics of \ivr systems have been refined since the beginning:
display and tracking properties, including sensors, technologies, and implementation, improving refresh rate, resolution, latency, and tracking fidelity.
\ivr systems can enhance users' sense of embodiment, presence, and appearance and behavior plausibility~\cite{Caserman2020SurveyFullBody, cummings2016immersive, wolf2022plausbilitydisplay} and have become essential in various fields.
They enable face-to-face-like interactions in social VR~\cite{smith2018communicationbehaviourvr, Young2015Dyadicinteractionsavatars}, aid therapy for body image disturbances \cite{dollinger2019vitras, portingale2024review} and physical rehabilitation \cite{Bartl2022EffectsAvatarEnvironment,Hepke2024DevelopmentValidation3D}, and improve motivation in training and exercise \cite{born2019exergamesimmersionembodiment}.
The effectiveness of these systems depends on how well they replicate the user's physical movement and how well they visualize the virtual environment \cite{cummings2016immersive}.
Research on \ivr systems focuses on technological advances and perceptual impacts.

Technological advances aim to enhance the visual fidelity, accuracy, and precision of these systems, thereby increasing their immersion.
Immersion refers to the objective technological capabilities of a system to deliver sensory stimuli, including display quality and tracking fidelity \cite{slater2003note,slater1997immersionandpresence}.
Improving the display properties can reduce negative effects like simulator sickness~\cite{wang2023framerateux}.
Increased tracking fidelity can lead to social interaction in VR closely resembling face-to-face communication behavior \cite{smith2018communicationbehaviourvr}.

Evaluations of the perceptual impacts of avatar embodiment and \ivr systems focus on presence and the sense of embodiment.
Presence is a subjective phenomenon often described as the sense of \enquote{being there} in a virtual environment~\cite{slater1997immersionandpresence}.
Research consistently demonstrates a positive correlation between immersion and presence.
Higher immersion (especially in the display characteristics) generally results in increased feelings of presence~\cite{Caserman2020SurveyFullBody,cummings2016immersive,slater2003note}.
The sense of embodiment refers to users' subjective feelings regarding owning and controlling virtual body representations~\cite{kilteni2012embodimentinvr}.
Kilteni et al.~\cite{kilteni2012embodimentinvr} defined the sense of embodiment with three dimensions: body ownership (the sensation of owning the virtual body), agency (the feeling of controlling the virtual body), and self-location (the sense of being situated within the virtual body).
Studies have shown that increased immersion positively influences the sense of embodiment, especially when improving visuomotor synchrony~\cite{Caserman2020SurveyFullBody,roth2020veq}.
Beyond these constructs, \ivr systems affect various other qualia, for example, spatial perception~\cite{Mohler2008fullbodyavatar}, communication and social presence \cite{merz2024voice, smith2018communicationbehaviourvr, Young2015Dyadicinteractionsavatars}, implicit attitudes~\cite{Peck2013Puttingyourselfskin}, and emotional responses~\cite{fiedler2024selfcues,waltemate2018impact,wolf2021embodiment}.
A related concept is appearance and behavior plausibility. It describes how authentic and congruent an avatar's appearance and behavior appear to users~\cite{mal2022virtual}.
Recent studies have demonstrated that higher immersion enhances appearance and behavior plausibility~\cite{wolf2022plausbilitydisplay}, thus improving user experience~\cite{mal2022virtual}.

\subsection{Hardware-Centered Evaluation of \ivr Systems}
\ivr systems vary greatly in how they capture and reconstruct human motion due to differences in hardware specifications, tracking methodologies, and software processing approaches.
Comparative studies highlighted substantial technical differences between these systems in terms of the tracking and reconstruction of motion quality (e.g., tracking accuracy, latency, and motion fidelity) \cite{Caserman2020SurveyFullBody, Hepke2024DevelopmentValidation3D, vox2021trackingcomparison} and display characteristics~\cite{wang2023framerateux}.
Marker-based systems generally offer higher precision, while inertial measurement unit (IMU) based or inside-out tracking approaches improve usability and setup simplicity at the cost of some accuracy \cite{Holzwarth2021ComparingAccuracyPrecision, vox2021trackingcomparison}.
Despite technical differences, the studies mentioned focus on objective system performance measurements, and objective metrics may not directly translate into user experience.
For example, less accurate \ivr systems may suffice in some contexts~\cite{Gutierrez2024ComparingOpticalCustom}, whereas accuracy and precision matter more in others~\cite{Merker2023MeasurementAccuracyHTC}.
Overall, this underscores that objective, hardware-centered metrics alone cannot fully explain subjective user experiences \cite{Gutierrez2024ComparingOpticalCustom, Merker2023MeasurementAccuracyHTC, yun2023animationfidelityselfavatar}.

\subsection{User-Centered Evaluation of \ivr Systems}
Subjective evaluations are essential for understanding how technological differences in \ivr systems affect user experience. 
Ganal et al.~\cite{ganal2020trackingsystemcomparison} proposed a framework that pairs objective measures (e.g., latency, positional and angular accuracy) with subjective reports (e.g., presence, body ownership), arguing that technical factors like latency can break visual congruence and trigger breaks in presence that negatively impact subjective experiences~\cite{Caserman2020SurveyFullBody, Kasahara2017MalleableEmbodimentChanging}.

Wang et al.~\cite{wang2023framerateux} found \unitframespersecond{120} to be a key threshold in user experience, with higher frame rates reducing simulator sickness and improving task performance, while frame rates under \unitframespersecond{60} force users to adopt compensatory strategies to cope with motion and detail loss.
Gon\c{c}alves et al.~\cite{goncalves2022impactbodytrackingsenseofembodiment} found that tracking fidelity, especially by tracking the hips, enhanced users' sense of embodiment.
Kimmel et al.~\cite{Kimmel2024KineticConnectionsExploring} showed that full-body tracking influenced user behavior and interpersonal spacing in social VR, but improved tracking fidelity, namely additional hand tracking, did not yield further measurable benefits in social presence.
Yun et al.~\cite{yun2023animationfidelityselfavatar} compared two \ivr systems based on sparse tracking and inverse kinematics (Unity IK, Final IK) and an IMU-based \ivr system with a high number of sensors (Xsens) in VR.
They evaluated the sense of embodiment when using the different systems and compared user performance across tasks, emphasizing body pose, lower-body, and upper-body interaction.
They found that Xsens achieved superior overall pose accuracy, but was outperformed by IK-based methods in tasks that require precise hand and foot placement due to latency issues and positional drifts of Xsens.
However, subjective measures of sense of embodiment were only captured after all tasks, limiting insights into how each task individually influenced user perception.
Lugrin et al. \cite{Lugrin2013Usabilitybenchmarksmotion} proposed benchmarks, combining objective performance (e.g., task success, error rates) with subjective experience measures, which was extended by related work \cite{ourbodytrackingcomparison}.

Overall, these findings highlight that technical improvements of \ivr systems, such as higher frame rate, tracking accuracy, positional fidelity, and reduced latency, do not uniformly or predictably improve user experiences.
Instead, subjective perceptions are context-dependent and influenced by the type of interactions users perform within the VR environment.
Consequently, user-centered comparisons are essential for identifying whether technical advancements improve user experience.

\subsection{Present Work}
Previous work has primarily focused on comparing contemporary, state-of-the-art \ivr systems, providing valuable insights into the capabilities of current technology.
However, these studies overlook how technological progress across system generations influences user experience.
To address this gap, we compare two generations of the same \ivr system to evaluate the evolution of technological improvements within a single system across multiple use cases. 
We employ a user-centered methodology to evaluate the following hypotheses using multiple tasks that simulate different use cases.
This approach allows us to understand how generational differences in full-body VR systems affect the user experience. 

Previous studies have shown a strong link between immersion and presence. They show that higher immersion generally leads to a greater sense of \enquote{being there} in a virtual environment, especially when improving display characteristics~\cite{cummings2016immersive, merz2024universalaccess, merz2024voice}, leading to this hypothesis:
\begin{itemize}
    \item \textbf{H1:} More immersive \ivr systems lead to an increased sense of presence.
\end{itemize}

Furthermore, prior work underscores the importance of immersion in enhancing the user's experience of the sense of embodiment, including body ownership and agency~\cite{born2019exergamesimmersionembodiment, merz2024universalaccess, merz2024voice}.
They show that this is caused by increased sensorimotor contingencies through higher tracking capabilities, resulting in this hypothesis:
\begin{itemize}
    \item \textbf{H2:} More immersive \ivr systems lead to an increased sense of embodiment.
\end{itemize}

In addition, research by Mal et al. \cite{mal2022virtual,mal2024vhpvr} indicates that appearance and behavior plausibility contribute to the user experience, and previous work showed that different immersive properties impact the perceived appearance and behavior plausibility \cite{wolf2022plausbilitydisplay}, which leads to the following hypothesis:
\begin{itemize}
    \item \textbf{H3:} More immersive \ivr systems lead to increased appearance and behavior plausibility.
\end{itemize}

Regarding task performance and task load, previous research suggests that technical advances in tracking fidelity, latency reduction, and display quality can decrease task load and influence task performance \cite{tong2023towards,wang2023framerateux,yun2023animationfidelityselfavatar}, leading to the following hypotheses: 
\begin{itemize}
    \item \textbf{H4:} More immersive \ivr systems lead to a reduced task load.
\end{itemize}
\begin{itemize}
    \item \textbf{H5:} More immersive \ivr systems lead to increased task performance.
\end{itemize}

%% file: sections/03-method.tex
\section{Method}

We conducted a user-centered, mixed study to evaluate how the two different generations of \ivr systems and their respective immersive characteristics influence user experience.
We used a $2\times5$ mixed design, with the \ivr system (old vs. new) as a between-subjects variable and VR task (five tasks) as a within-subjects variable.
The data of the new \ivr system is also presented in a related study, which describes the tasks in more detail \cite{ourbodytrackingcomparison}.
The choice of the \ivr system as a between-condition was guided by the following reasons: repeating five tasks with two systems would have doubled session duration, increasing fatigue, reducing engagement, and introducing learning effects. To balance rigor with practicality, we therefore treated the \ivr system as a between-subjects factor, while still capturing diverse use cases through the within-subjects task variation.
Therefore, participants were randomly assigned to either the old or new system condition and completed five VR tasks in counterbalanced order.
We measured presence, sense of embodiment, appearance and behavior plausibility, task load, and objective performance metrics for each task.
The study complied with the Declaration of Helsinki and was approved by the ethics committee of the Institute for Human-Machine-Media (MCM) at the University of Würzburg.

\subsection{System Description}
\label{sec:systemdescription}

\begin{figure}[!t]
    \centering
    \includegraphics[width=\columnwidth,trim={0mm 0mm 0mm 0mm},clip]{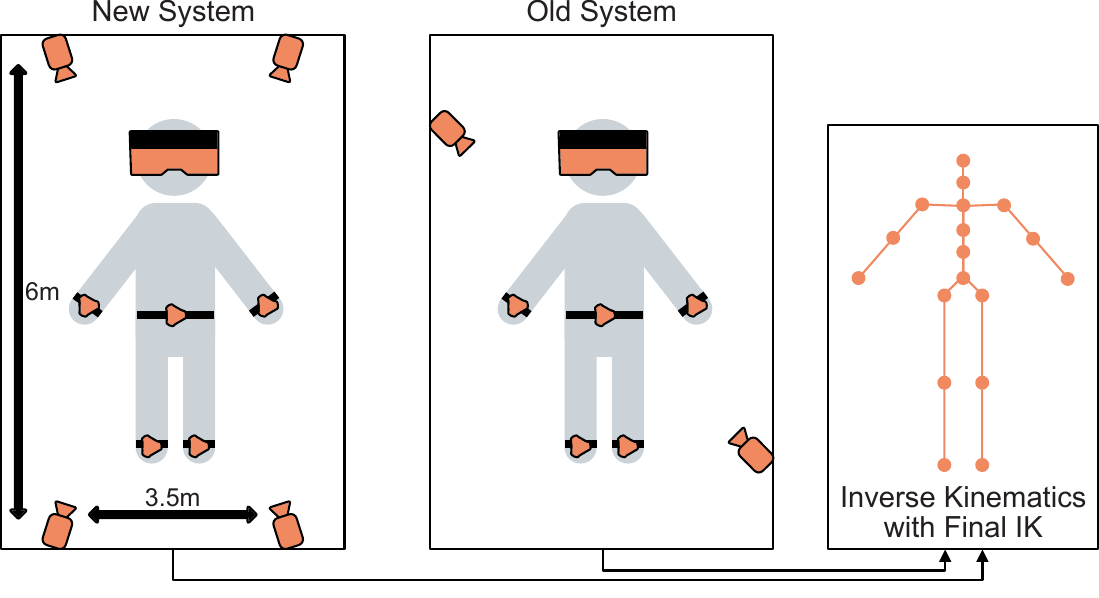}
    \caption{The figure shows (1) the VR hardware components of the new (left) and the old \ivr system (middle), and (2) the avatar's pose generation from the hardware's data (right).} 
    \label{fig:tracking-system-components}
\end{figure}

We compared two \ivr systems from different hardware generations but with similar core configurations: HMD, 6-point tracking via body-worn trackers, a VR-capable gaming PC, and inverse kinematics (IK) for full-body pose reconstruction.
A direct comparison is shown in \autoref{tab:comparison-fba-systems}.
We chose the approach of body-worn sparse trackers and inverse kinematics as this system is widely used in recent research \cite{ourbodytrackingcomparison}.
The old \ivr system used an HTC Vive HMD, providing a resolution of $1080 \times 1200$ pixels per eye, a \unitdegree{110} horizontal field of view, and a refresh rate of \unithertz{90}.
The headset weighs \unitgram{470}.
For tracking, the system used two first-generation SteamVR Base Stations and five HTC Vive Tracker 1.0 (\unitgram{90} each), offering coverage for a play area of $3.5 \times 3.5$ meters.
The base stations operated at \unithertz{60} with a field of view of $120^\circ\times120^\circ$.
In contrast, the new \ivr system used an HTC Vive Pro 2, with a resolution of $2448 \times 2448$ pixels per eye, a \unitdegree{120} horizontal field of view, and a \unithertz{120} refresh rate.
The headset weighs \unitgram{850}.
Since the second generation SteamVR Base stations support more than two, we used four base stations, leveraging a better coverage paired with five HTC Vive Tracker 3.0 (\unitgram{75} each).
This setup enabled a play area of approximately $7 \times 7$ meters, with tracking performed at \unithertz{100} and base stations offering a \unitdegree{160} horizontal and \unitdegree{120} vertical field of view.
The old or new trackers were attached to the participants using velcro straps: on the back of their hands, on the top of their feet, and around the lower back using a belt as shown in \autoref{fig:tracking-system-components}.
Both systems were powered by the same VR-capable gaming PC running Windows 10 and equipped with an Intel Core i9-10900K processor, NVIDIA GeForce RTX 3090 GPU, and 64~GB of RAM.
Overall, the new \ivr system offers higher immersion with improved display characteristics (higher resolution and refresh rate), expanded tracking coverage, and reduced motion-to-photon latency, but a higher weight of the HMD.

\begin{table}[t]
\caption{Comparison of the \ivr systems used in our study.}
\label{tab:comparison-fba-systems}
\begin{tabular*}{\columnwidth}{@{\extracolsep{\fill}}p{0.25\columnwidth}p{0.375\columnwidth}p{0.375\columnwidth}@{}}
\toprule
 & \textbf{Old System} & \textbf{New System} \\
\midrule
 & \textbf{HTC Vive} & \textbf{HTC Vive Pro 2} \\
\textbf{Resolution per eye} & $1080 \times 1200$ & $2448 \times 2448$ \\
\textbf{Field of View} & \unitdegree{110} & \unitdegree{120} \\
\textbf{Refresh Rate} & \unithertz{90} & \unithertz{120} \\
\textbf{Weight} & \unitgram{470} & \unitgram{850} \\
\midrule
 & \textbf{Base Stations 1.0} & \textbf{Base Stations 2.0} \\
\textbf{Count} & 2 & 4 \\
\textbf{Play Area} & $3.5 \times 3.5$ meters & $7 \times 7$ meters \\
\textbf{Refresh Rate} & \unithertz{60} & \unithertz{100} \\
\textbf{Field of View} & $120^\circ\times120^\circ$ & $160^\circ\times120^\circ$ \\
\midrule
 & \textbf{Vive Tracker 1.0} & \textbf{Vive Tracker 3.0} \\
\textbf{Count} & 5 & 5 \\
\textbf{Weight} & \unitgram{90} & \unitgram{75} \\
\midrule
 & \textbf{Overall} & \textbf{Overall} \\
\textbf{Motion-to-photon latency} & \multirow{2}{*}{\unitmillisecond{85.94}} & \multirow{2}{*}{\unitmillisecond{16.15}} \\
\bottomrule
\end{tabular*}
\end{table}

We developed the VR application using Unity 2022.03.52f1~LTS and integrated the HMD and the tracker using SteamVR~2.9 and an abstract layer for VR device integration \cite{kern2023reality}.
To accurately map tracking positions and rotations to specific body parts, we employed a custom calibration algorithm that requires users to adopt a standard T-pose (standing upright with arms extended horizontally). 
The tracker transforms were systematically assigned to body joints from top to bottom and left to right.
Participants embodied VALID avatars selected based on self-reported gender and ethnicity to ensure accurate representation \cite{do2023valid}. 
Avatar animations were generated through the character processor architecture \cite{merz2024pipelining}, relying on positional and rotational data from tracked devices.
Human poses were inferred using inverse kinematics provided by FinalIK~2.2.0.

We measured the motion-to-photon latency of the body pose for both \ivr systems by counting frames between real and rendered movements~\cite{stauffert2020latencyreview}.
Using two iPhone 13 high-speed cameras, we recorded the user's motions and the corresponding avatars' reactions through the display of the HMD at \unitframespersecond{240}. The average latency of the body pose for the new system was \unitmillisecond{16.15} ($SD=2.50$), while it was \unitmillisecond{85.94} ($SD=22.95$) for the old system.

\subsection{Tasks}
\label{sec:tasks}
Participants completed five VR tasks, designed to simulate different use cases, each targeting specific aspects of \ivr systems, including pose realism (Mirror Task), tracking latency (Hallway and Wire Task), and positional accuracy (Dance and Reaction Wall Task).
Detailed descriptions of these tasks can be found in the comparison of different approaches of VR systems for full-body avatars by Tschanter et al., which is currently under review \cite{ourbodytrackingcomparison}.
\autoref{fig:teaser} illustrates the first-person perspective of each task. 
Each task lasted about 120 seconds, and the order of completion was randomized to minimize order effects.

\subsection{Measures}
\label{sec:measures}

Participants completed questionnaires using a self-hosted instance of LimeSurvey~4\footnote{https://www.limesurvey.org}.
We ensured questionnaire accuracy by using validated German translations or conducting back-and-forth translations. 
We assessed subjective user experience through questionnaires covering presence, sense of embodiment, appearance and behavior plausibility, and task load (listed in \autoref{tab:quesionnaires}).
As control measures, we captured signs of simulator sickness and the participant's immersive tendencies.
We collected objective performance metrics for each of the five tasks.
These measures captured aspects such as accuracy, precision, and reaction time.
A detailed overview of task-specific metrics is provided in \autoref{tab:taskperformancemeasures}.
We gathered qualitative feedback via open-text questions in LimeSurvey after each task and at the end of the study.
Therefore, participants answered a question targeting the comfort of the \ivr system:
\enquote{How comfortable do you find the VR system (including all equipment) that you wore during the task/tasks?}; and a question targeting the perceived task difficulty: \enquote{How difficult or easy did you find the task/s?}

\begin{table}[t]
    \caption{Overview of the questionnaires used during the study.}
    \label{tab:quesionnaires}
    \begin{tabular*}{\columnwidth}{@{\extracolsep{\fill}}p{0.25\columnwidth}p{0.4\columnwidth}p{0.14\columnwidth}@{}}
        \toprule
        \textbf{Questionnaire} & \textbf{Measure} & \textbf{Range} \\ 
        \midrule
        \multicolumn{3}{l}{\hspace{-2mm}\textbf{Presence}} \\
        \hspace{2mm}IPQ~\cite{schubert2001experience} & General Presence & [0--6] \\
        & Spatial Presence & [0--6] \\
        & Involvement & [0--6] \\
        \multicolumn{3}{l}{\hspace{-2mm}\textbf{Sense of Embodiment}} \\
        \hspace{2mm}VEQ~\cite{roth2020veq} & Virtual Body Ownership & [1--7] \\
        & Agency & [1--7] \\
        \hspace{2mm}VEQ+~\cite{fiedler2023embodiment} &  Self-Location & [1--7] \\
        \multicolumn{3}{l}{\hspace{-2mm}\textbf{Appearance and Behavior Plausibility}} \\
        \hspace{2mm}VHP~\cite{mal2022virtual,mal2024vhpvr} & Appearance and Behavior Plausibility & [1--7] \\
        \multicolumn{3}{l}{\hspace{-2mm}\textbf{Task Load}} \\
        \hspace{2mm}RTLX~\cite{hart2006rawtlx} & Overall Task Load & [0--100] \\
        & Mental Demand & [0--100] \\
        & Physical Demand & [0--100] \\
        & Frustration & [0--100] \\
        & Temporal Demand & [0--100] \\
        & Performance & [0--100] \\
        & Effort & [0--100] \\
        \multicolumn{3}{l}{\hspace{-2mm}\textbf{Control Measures}} \\ 
        \hspace{2mm}VRSQ~\cite{kim2018vrsq} & Simulator Sickness &  [0--100] \\ 
        \hspace{2mm}ITQ~\cite{witmer1994measuring} & Immersive Tendencies &  [1--7] \\
        \bottomrule
    \end{tabular*}
\end{table}

\begin{table}[t]
    \caption{Overview of the objective measurements for task performance used during the study.}
    \label{tab:taskperformancemeasures}
    \begin{tabular*}{\columnwidth}{@{\extracolsep{\fill}}p{0.31\columnwidth}p{0.8\columnwidth}@{}}
        \toprule
        \textbf{Task} & \textbf{Performance Measure} \\ 
        \midrule
        \textbf{Mirror Task} & -- \\
        \textbf{Hallway Task} 
        & Number of walls passed \\
        & Number of walls hit \\
        \textbf{Wire Task} 
        & Wire distance in meters\\
        & Number of times the wire was left  \\
        \textbf{Reaction Wall Task} 
        & Reaction time in seconds\\
        & Number of correct touched targets \\
        \textbf{Dance Task} 
        & Reaction time in seconds\\
        & Number of correct touched tiles \\
        \bottomrule
    \end{tabular*}
\end{table}

\subsection{Procedure}
\label{sec:procedure}

\begin{figure}[!t]
    \centering\includegraphics[width=\columnwidth,trim={0mm 0mm 0mm 0mm},clip]{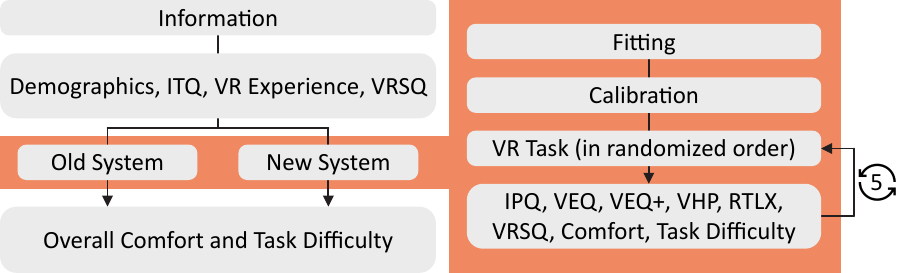}
    \caption{The figure shows the study procedure with the VR exposure on the right. Participants underwent five VR tasks in randomized order, either in the new \ivr system or the old one.} 
    \label{fig:procedure}
\end{figure}

Our study followed a standardized procedure, outlined in \autoref{fig:procedure}.
The average duration of the study was \unitminute{68}, with each VR session lasting about \unitminute{2.5}.
Participants first read the participant information and gave their written informed consent.
They then completed pre-questionnaires, including demographic information, ITQ, and VRSQ.
Before starting the VR tasks, participants received instructions and help in putting on the HMD and body trackers.
At the start of each task, we calibrated the avatar and the \ivr system.
A virtual whiteboard showed the written task instructions and the participants initiated each task by pressing a virtual button.
Following the task, participants completed post-experience questionnaires.
This process was repeated until all five tasks were completed.
Finally, participants were required to complete post-questionnaires.

\subsection{Participants}
\label{sec:participants}

To determine our sample size, we calculated a priori power analysis for a repeated-measures ANOVA with between-subject factors, which indicated a required sample size of $N = 40$. 
For this analysis, we assumed a small-to-medium effect size of $f = 0.222$~\cite{cohen1988statistical}, informed by similar studies \cite{eubanks2020bodytrackingfidelityik, yun2023animationfidelityselfavatar}, an alpha level of $0.05$, a statistical power of $0.80$, and a within-subject correlation of $r = 0.2$, anticipating relatively low within-subject correlation.
We recruited participants via the local participant management system. 
They were either undergraduate participants receiving course credit or participants receiving monetary compensation.
Inclusion criteria were: (1) normal or corrected vision and hearing, (2) at least ten years of German language proficiency, and (3) no reported sensitivity to simulator sickness.

After we collected 40 participants, the sample showed a substantial age imbalance ($U(20,20) = 3700$, $p < .003$), calculated using a Mann-Whitney U test due to non-normality of the data.
The participants of the old \ivr system had a mean age of 24.65 years ($SD = 4.13$), whereas those using the new system averaged 32.43 years ($SD = 4.13$), with overall seven participants above the age of 35.
To address this, we initially attempted to recruit additional participants over the age of 35 to achieve an age-balanced sample.
However, recruitment constraints made this unfeasible. We therefore restricted the sample to participants under the age of 35 years, which yielded an age-balanced dataset across conditions.
Consequently, we continued data collection to reach 40 participants under 35 years old, ensuring that differences could be attributed more confidently to system generation rather than demographic variation.

Therefore, the final age-balanced sample included 40 participants (30 female, 10 male), aged 19–34 years ($M = 24.05$, $SD = 3.95$). No significant group differences were found between both \ivr systems in simulator sickness ($U(20,20) = 177$, $p = .540$), immersive tendencies ($U(20,20) = 225$, $p = .507$), prior VR experience, or age ($U(20,20) = 218$, $p = .633$), with mean ages of 24.25 ($SD = 4.04$) for the old and 23.85 ($SD = 4.03$) for the new \ivr system.

%% file: sections/04-result.tex
\section{Results}
In the first subsection, we investigate the effects of \ivr systems and tasks on participants' presence, sense of embodiment, appearance and behavior plausibility, task load, and task performance.
In the second subsection, we analyze the qualitative data collected. 
All statistical analyses were conducted using R version 2024.12.01\footnote{https://www.r-project.org/}.

\subsection{Quantitative Analysis}
\begin{figure*}[!t]
    \centering
    \begin{subfigure}{0.2\textwidth}
        \includegraphics[width=\linewidth]{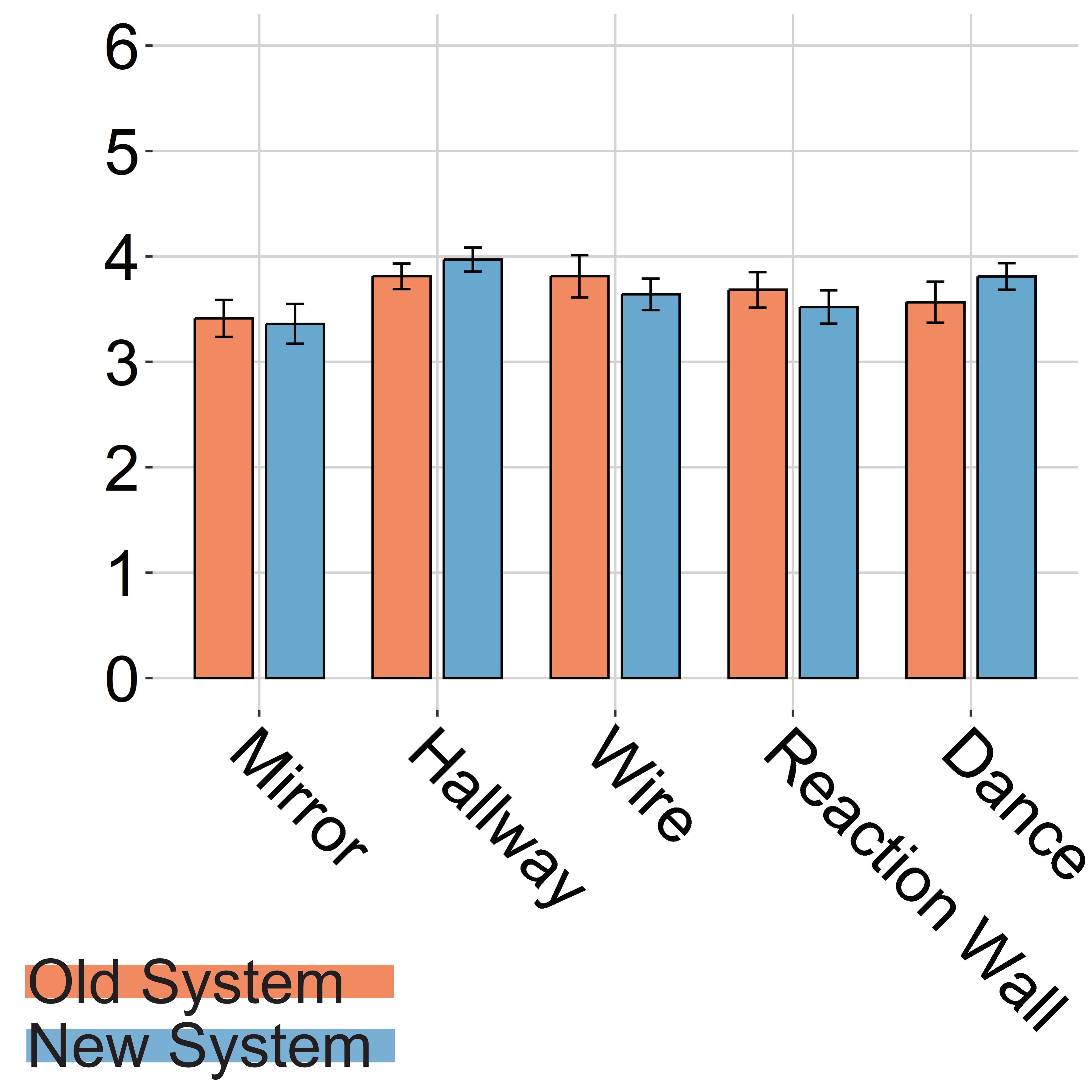}
        \caption{Spatial Presence}
        \label{fig:spatial-presence}
    \end{subfigure}\hfill
    \begin{subfigure}{0.2\textwidth}
        \includegraphics[width=\linewidth]{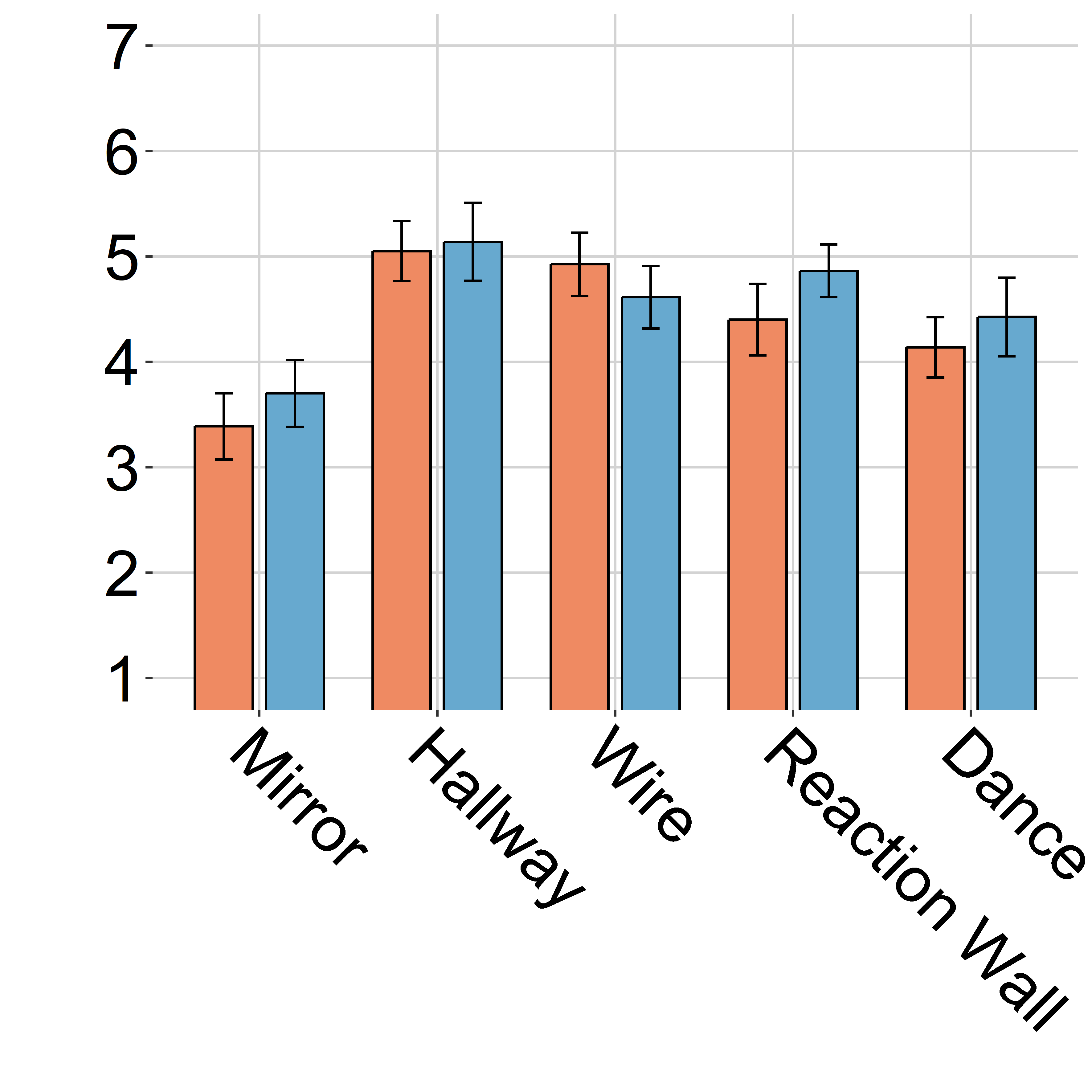}
        \caption{Body Ownership}
        \label{fig:body-ownership}
    \end{subfigure}\hfill
    \begin{subfigure}{0.2\textwidth}
        \includegraphics[width=\linewidth]{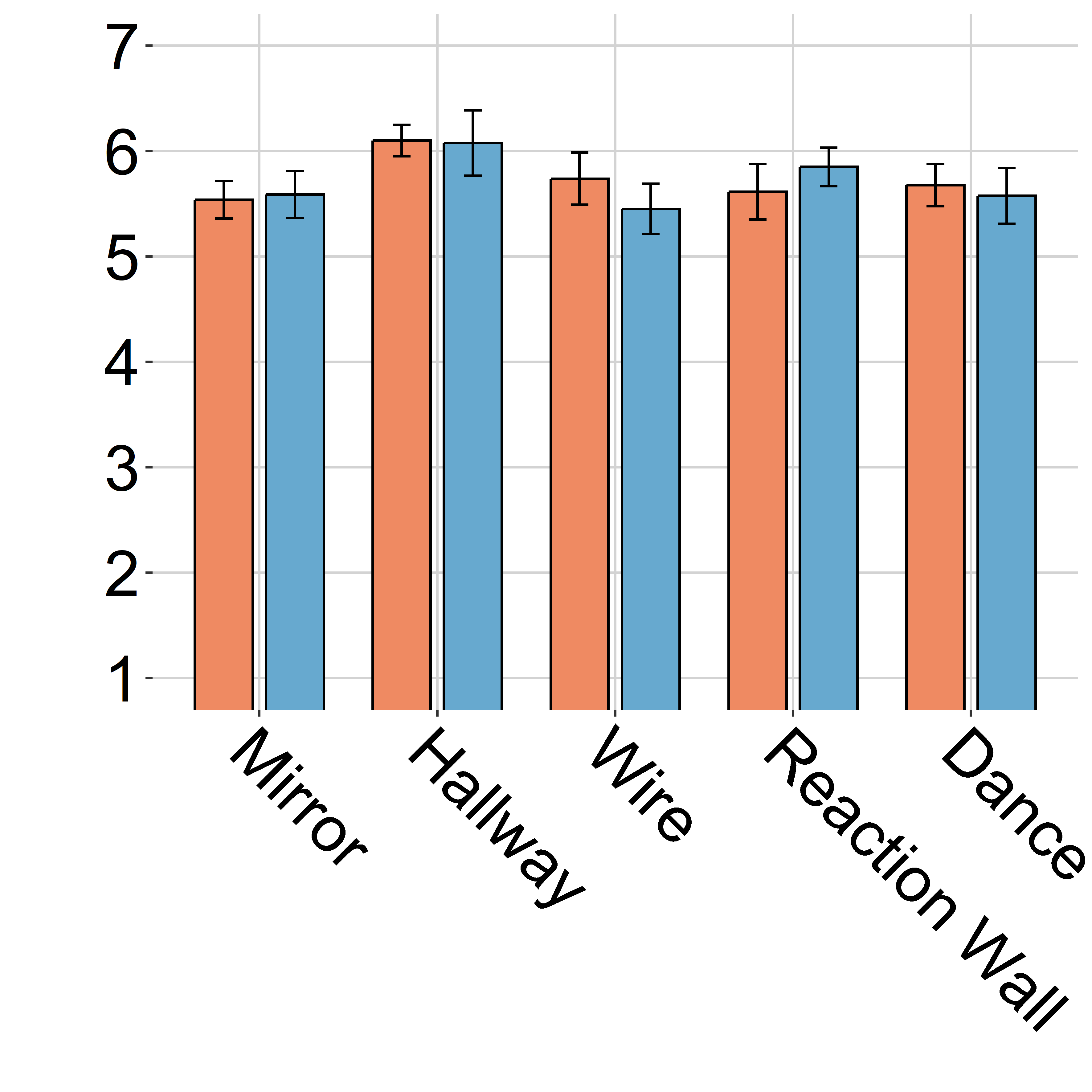}
        \caption{Agency}
        \label{fig:agency}
    \end{subfigure}\hfill
    \begin{subfigure}{0.2\textwidth}
        \includegraphics[width=\linewidth]{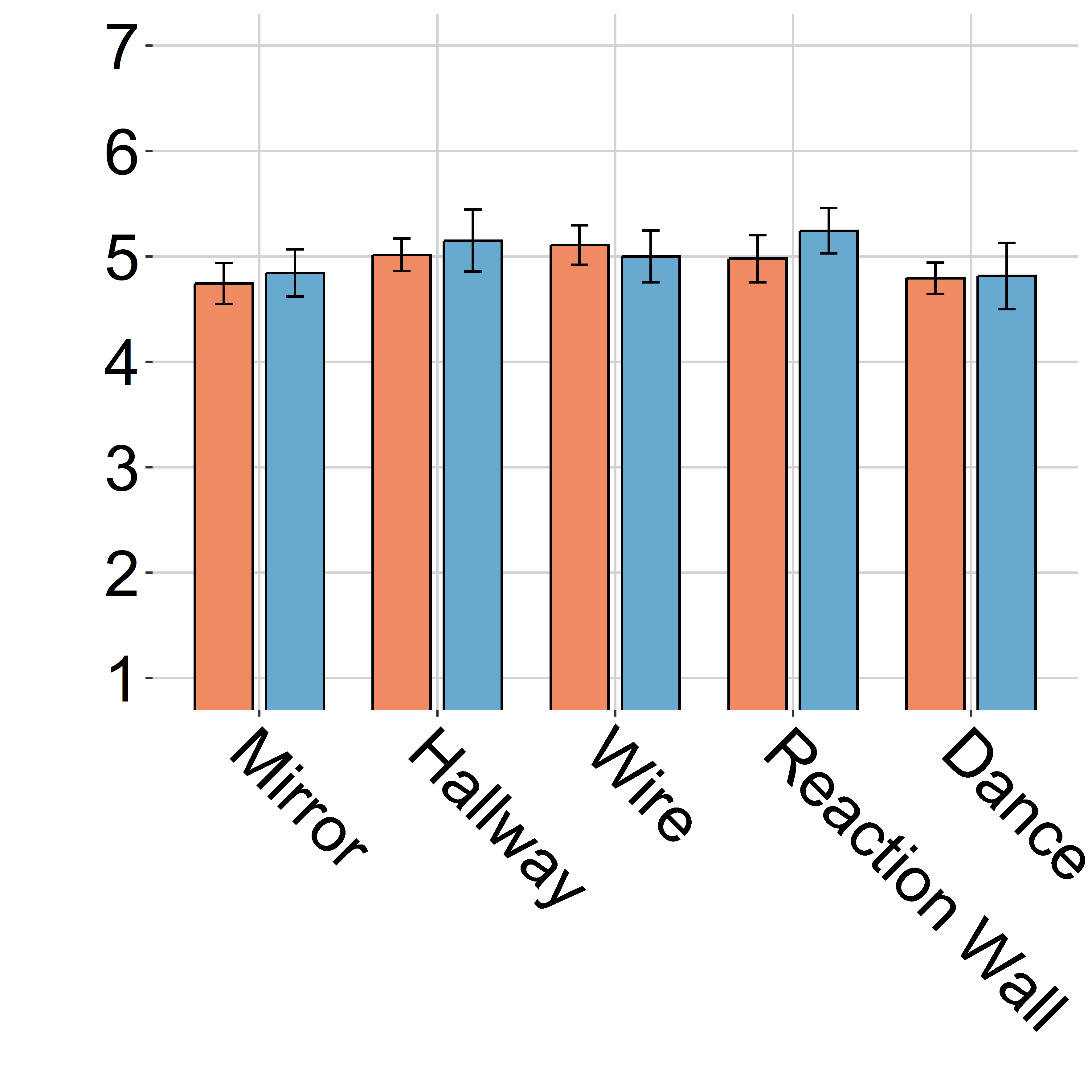}
        \caption{App. and Behavior Plausibility}
        \label{fig:abp}
    \end{subfigure}\hfill
    \begin{subfigure}{0.2\textwidth}
        \includegraphics[width=\linewidth]{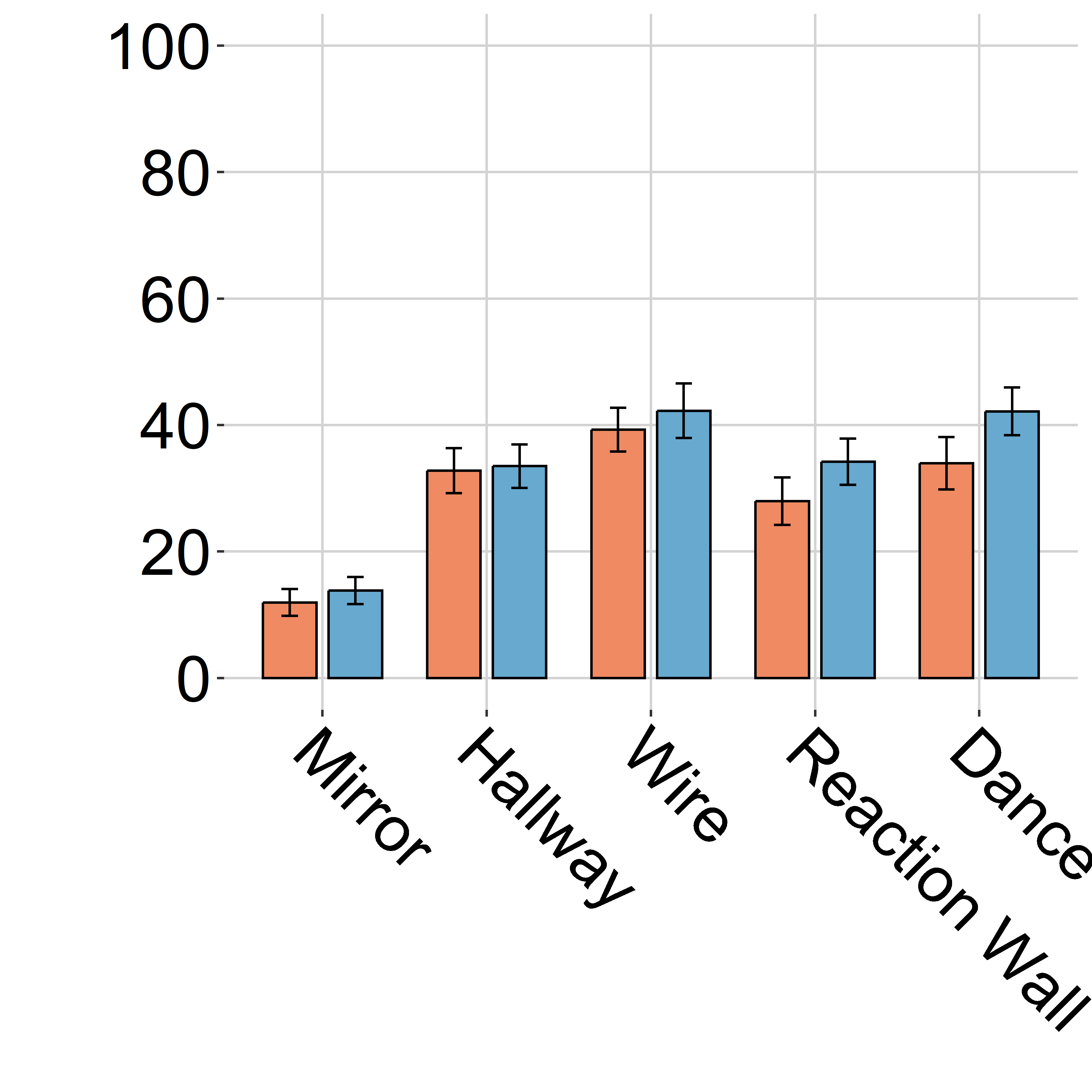}
        \caption{General Task Load}
        \label{fig:taskload}
    \end{subfigure}
    \caption{These figures outline interaction plots for selected mean values for the user experience qualia. The task factor is plotted on the x-axis, while the separate lines represent the \ivr system factor. Error bars show the standard error.}
    \label{fig:means-ux-qualia}
\end{figure*}

We performed $2\times5$ mixed ANOVAs for each variable of presence, sense of embodiment, appearance and behavior plausibility, and task load.
Due to violations of normality and homoscedasticity, we applied a robust variation with the trimmed means ANOVA with a 20\% cut-off using the R package WRS2 \cite{mair2020wrs2}. 
Following a significant main effect of the task, we ran post hoc Wilcoxon signed-rank tests with Bonferroni alpha correction.
To compare task performance between the two \ivr systems for each task, we used Mann-Whitney U tests, since the assumptions of normality and homoscedasticity were also not met. 
For all measurements, where hypotheses were not supported, we performed Bayesian ANOVAs using Bayes Factors from the BayesFactor package~\cite{morey2018bayesfactor}, comparing each model against a null model with subject ID as a random effect.
All tests were performed with $\upalpha = .05$. 
Descriptive results and p-values for the user experience measures are shown in \autoref{tab:descriptive-results}, while selected measures are visualized in \autoref{fig:means-ux-qualia} and descriptive measures of the task performance measures are shown in \autoref{tab:descriptive-results-performance}.
For clarity, we report only significant test statistics of the mixed ANOVAs and Mann-Whitney U tests in the text below.

\begin{table*}[hbt]
    \caption{The table shows descriptive values for each experimental condition and p-values of the main and interaction effects for the factors \ivr system and task for the trimmed means mixed ANOVA. Statistical significance indicators: ${^{*}\,p<0.05}$; ${^{\dag}\,p< 0.01}$; ${^{\ddag}\,p< 0.001}$.}
    \scriptsize
    \label{tab:descriptive-results}
    \adjustbox{max width=\textwidth}{
    \begin{tabular}{lcccccc|ccc}
    \toprule
    &  & Mirror & Hallway & Wire & Reaction Wall & Dance & Main & Main & Interaction \\
    \cmidrule{3-7}
    & \ivr System & $M$ ($SD$) & $M$ ($SD$) & $M$ ($SD$) &  $M$ ($SD$) & $M$ ($SD$) & \ivr System & Task &  \\
    \midrule

    \multicolumn{4}{l}{\textbf{Presence}} \\
    \hspace{4mm}\multirow{2}{*}{Spatial} & Old & 3.41 (0.73) & 3.81 (0.5) & 3.81 (0.83) & 3.68 (0.7) & 3.56 (0.8) & $p = .799$ & $p = .010^{*}$ & $p = .181$ \\
    & New & 3.36 (0.85) & 3.97 (0.51) & 3.64 (0.67) & 3.52 (0.7) & 3.81 (0.56)  \\
    \cmidrule{2-7}
    \hspace{4mm}\multirow{2}{*}{Involvement} & Old & 2.75 (0.92) & 3.21 (0.81) & 3.36 (0.59) & 3.30 (0.57) & 3.28 (0.78) & $p = .410$ & $p = .427$ & $p = .181$ \\
    & New & 3.22 (0.90) & 3.22 (0.65) & 3.22 (0.67) & 3.22 (0.69) & 3.40 (0.72)   \\
    \cmidrule{2-7}
    \hspace{4mm}\multirow{2}{*}{General} & Old & 3.55 (1.23) & 4.30 (1.17) & 4.35 (1.04) & 3.95 (1.31) & 4.00 (0.79) & $p = .783$ & $p = .002^{\dag}$ & $p = .942$ \\
    & New & 3.60 (1.27) & 4.55 (1.15) & 4.05 (1.36) & 4.10 (1.17) & 4.10 (1.21)  \\

    \addlinespace
    \multicolumn{4}{l}{\textbf{Sense of Embodiment}} \\
    \hspace{4mm}\multirow{2}{*}{VBO} & Old & 3.39 (1.41) & 5.05 (1.27) & 4.92 (1.35) & 4.40 (1.52) & 4.14 (1.28) & $p = .399$ & $p < .001^{\ddag}$ & $p = .421$ \\
    & New & 3.70 (1.41) & 5.14 (1.65) & 4.61 (1.32) & 4.86 (1.12) & 4.42 (1.67)  \\
    \cmidrule{2-7}
    \hspace{4mm}\multirow{2}{*}{Agency} & Old & 5.54 (0.79) & 6.10 (0.67) & 5.74 (1.10) & 5.61 (1.18) & 5.68 (0.90) & $p = .712$ & $p = .003^{\dag}$ & $p = .311$ \\
    & New & 5.59 (1.00) & 6.08 (1.39) & 5.45 (1.07) & 5.85 (0.82) & 5.58 (1.18)  \\
    \cmidrule{2-7}
    \hspace{4mm}\multirow{2}{*}{Self-Location} & Old & 3.55 (1.27) & 4.53 (1.10) & 4.28 (1.29) & 4.11 (1.03) & 4.10 (0.95) & $p = .934$ & $p = .012^{*}$ & $p = .450$ \\
    & New & 3.44 (1.45) & 4.22 (1.35) & 4.38 (1.34) & 4.16 (1.07) & 4.24 (1.41)  \\

    \addlinespace
    \multicolumn{4}{l}{\textbf{Appearance and Behavior Plausibility}} \\
    \hspace{4mm}\multirow{2}{*}{} & Old & 4.74 (0.87) & 5.01 (0.69) & 5.11 (0.84) & 4.98 (1.00) & 4.79 (0.67) & $p = .541$ & $p = .078$ & $p = .684$ \\
    & New & 4.84 (1.00) & 5.15 (1.32) & 5.00 (1.10) & 5.24 (0.96) & 4.81 (1.41)  \\

    \addlinespace
    \multicolumn{4}{l}{\textbf{Task Load}} \\
    \hspace{4mm}\multirow{2}{*}{General} & Old & 11.96 (9.47) & 32.79 (15.99) & 39.29 (15.45) & 27.96 (16.73) & 33.96 (18.60) & $p = .616$ & $p < .001^{\ddag}$ & $p = .767$ \\
    & New & 13.83 (9.55) & 33.50 (15.35) & 42.25 (19.29) & 34.21 (16.41) & 42.17 (16.94) \\
    \cmidrule{2-7}

    \hspace{4mm}Mental & Old & 11.75 (14.62) & 20.50 (15.21) & 46.50 (31.71) & 24.00 (24.04) & 32.00 (23.02) & $p = .947$ & $p < .001^{\ddag}$ & $p = .963$ \\
    \hspace{4mm}Demand & New & 15.00 (20.90) & 19.00 (11.31) & 48.50 (27.39) & 27.25 (22.85) & 33.00 (23.47)  \\
    \cmidrule{2-7}
    \hspace{4mm}Physical & Old & 18.25 (17.27) & 48.75 (26.40) & 27.00 (21.48) & 34.25 (23.69) & 42.00 (23.14) & $p = .770$ & $p < .001^{\ddag}$ & $p = .637$ \\
    \hspace{4mm}Demand & New & 21.25 (21.08) & 48.25 (24.78) & 29.25 (25.92) & 38.75 (22.99) & 55.75 (22.20)  \\
    \cmidrule{2-7}
    \hspace{4mm}\multirow{2}{*}{Frustration} & Old & 13.25 (16.49) & 13.00 (15.68) & 43.75 (26.00) & 16.00 (17.06) & 27.00 (25.41) & $p = .759$ & $p < .001^{\ddag}$ & $p = .608$ \\
    & New & 7.25 (8.19) & 15.50 (16.61) & 44.25 (26.77) & 16.25 (17.76) & 29.50 (21.21)  \\
    \cmidrule{2-7}  
    \hspace{4mm}Temporal & Old & 8.00 (12.92) & 45.75 (27.83) & 29.25 (26.57) & 36.25 (29.69) & 31.50 (28.57) & $p = .342$ & $p < .001^{\ddag}$ & $p = .453$ \\
    \hspace{4mm}Demand & New & 6.00 (3.84) & 46.00 (31.90) & 27.00 (23.02) & 52.50 (31.69) & 46.50 (28.24)  \\
    \cmidrule{2-7}
    \hspace{4mm}\multirow{2}{*}{Performance} & Old & 8.50 (8.29) & 29.50 (26.45) & 50.75 (21.42) & 19.25 (16.16) & 32.25 (22.68) & $p = .089$ & $p < .001^{\ddag}$ & $p = .872$ \\
    & New & 26.00 (25.73) & 31.25 (23.33) & 63.00 (17.35) & 33.50 (24.71) & 39.75 (19.83) \\
    \cmidrule{2-7}
    \hspace{4mm}\multirow{2}{*}{Effort} & Old & 12.00 (10.81) & 39.25 (25.87) & 38.50 (27.10) & 38.00 (27.07) & 39.00 (22.04) & $p = .952$ & $p < .001^{\ddag}$ & $p = .694$ \\
    & New & 7.50 (7.34) & 41.00 (22.75) & 41.50 (27.73) & 37.00 (24.52) & 48.50 (26.16) \\
    \bottomrule
    \end{tabular}}
\end{table*}

\subsubsection{Presence}
For spatial presence, we found a main effect of the task ($Q(4,15.25) = 4.82, p = .010$).
Post-hoc comparisons indicated significantly higher scores in the hallway compared to the mirror task ($W(40) = 440.5$, $p = .010$).
We found no effect for the \ivr system and no interaction effect. Since H1 was not confirmed, we calculated a Bayesian ANOVA indicating evidence for no effect of the \ivr system ($BF_{01} = 2.646$).

No significant effects were found for involvement for the \ivr system, task, or interaction. Since H1 was not confirmed, we calculated a Bayesian ANOVA indicating moderate evidence for no effect of the \ivr system ($BF_{01} = 3.145$).

For general presence, we observed a significant main effect of the task ($Q(4,16.91) = 6.92, p = .002$). Post-hoc comparisons indicated significantly higher scores in the hallway compared to the mirror task ($W(40) = 264$, $p < .001$). The \ivr system and interaction effects were not significant. Bayesian ANOVA supports the null findings for H1 with $BF_{01} = 2.994$ for the \ivr system.

\subsubsection{Sense of Embodiment}
For virtual body ownership, there was a significant main effect of the task ($Q(4,17.15) = 8.360, p > .001$). Post-hoc tests revealed significantly lower scores for the mirror task vs. hallway ($W(40) = 644.5$, $p < .001$), vs. wire ($W(40) = 70.5$, $p < .001$), and vs. reaction wall ($W(40) = 104.5$, $p = .006$). There were also significantly lower values for the dance task compared to the hallway task ($W(40) = 150.5$, $p = .023$). We found no effects of the \ivr system or interaction. Since H2 was not confirmed, there is moderate evidence that the null hypothesis is more likely ($BF_{01} = 3.268$).

For agency, the task had a significant effect ($Q(4,16.74) = 6.17, p = .003$). Post-hoc comparisons indicated higher agency for the hallway compared to the mirror task ($W(40) = 500$, $p = .005$), the dance task ($W(40) = 126$, $p = .033$), and the wire task ($W(40) = 515$, $p = .042$). We found no significant effects for the \ivr system or its interaction. Bayesian ANOVA supports the null findings for H2 for the \ivr system ($BF_{01} = 3.334$).

For self-location, we found a significant main effect of the task ($Q(4,17.04) = 4.42, p = .012$). Post-hoc comparisons showed significantly higher scores for dance vs. mirror ($W(40) = 520$, $p = .008$), hallway vs. mirror ($W(40) = 638$, $p = .001$), wire vs. mirror ($W(40) = 92.5$, $p < .001$), and reaction wall vs. mirror ($W(40) = 122$, $p = .009$). We found no effects of the \ivr system or interaction. Bayesian ANOVA supports the null findings of H2 with $BF_{01} = 2.653$ for the \ivr system.

\subsubsection{Appearance and Behavior Plausibility}
We observed no main or interaction effect for appearance and behavior plausibility. Bayesian ANOVA supports the null findings of H2 with $BF_{01} = 2.959$) for the \ivr system.

\subsubsection{Task Load}
The analysis of the general task load showed a significant main effect of the task ($Q(4,16.79) = 24.09$, $p < .001$). The significant post-hoc comparisons of this and the following main effects of the task load variables can be found in \autoref{tab:wilcoxon_taskload1}.
No significant effect of the \ivr system or interaction was found. Bayesian ANOVA supports the null findings for H4 with $BF_{01} = 2.451$ for the \ivr system.

For mental demand, we found a significant main effect of the task ($Q(4,16.87) = 11.17$, $p < .001$). 
No effects of the \ivr system or interaction were observed. Since H4 was not confirmed, we calculated a Bayesian ANOVA indicating moderate evidence for no effect of the \ivr system ($BF_{01} = 3.610$).

Physical demand showed a significant effect of the task ($Q(4,16.80) = 23.14$, $p < .001$). 
No effects of the \ivr system or interaction were observed. Bayesian ANOVA supports the null findings for H4 for the \ivr system ($BF_{01} = 2.685$).

Frustration ratings showed a significant task effect ($Q(4,16.85) = 11.55$, $p < .001$). We found no effects of the \ivr system or interaction. Since H4 was not confirmed, we calculated a Bayesian ANOVA indicating moderate evidence for no effect of the \ivr system ($BF_{01} = 4.673$).

Temporal demand revealed a significant task effect ($Q(4,16.32) = 11.10$, $p < .001$),  but no significant \ivr system or interaction effect.
Bayesian ANOVA supports the null findings for H4 for the \ivr system ($BF_{01} = 2.967$).

For performance, there was a significant main effect of the task ($Q(4,16.44) = 19.54$, $p < .001$), but no significant effect of the \ivr system or interaction effect.
Bayesian ANOVA weakly supports the null findings for H4 for the \ivr system ($BF_{01} = 0.581$).

Finally, the perceived effort yielded a significant main effect of the task ($Q(4,17.60) = 17.59$, $p < .001$). There was no main effect of the \ivr system or interaction. Since H4 was not confirmed, we calculated a Bayesian ANOVA indicating moderate evidence for no effect of the \ivr system ($BF_{01} = 3.745$).

\subsubsection{Task Performance}
For task performance measures, one participant was excluded from the hallway task and three from the wire task due to technical issues during data collection. 
No significant performance differences were found between the old and new \ivr systems for any task, as shown in \autoref{tab:wilcoxon_tests}. 
Since H5 was not confirmed, we calculated Bayesian ANOVAs which support the null findings for the \ivr system for the hallway (number of walls passed: $BF_{01} = 3.165$; number of walls hit: $BF_{01} = 3.145$), wire (wire distance in meters: $BF_{01} = 2.890$; number of times the wire was left: $BF_{01} = 3.125$, reaction wall (reaction time in seconds: $BF_{01} = 2.874$; number of correct touched targets: $BF_{01} = 3.215$) and the dance task (reaction time in seconds: $BF_{01} = 2.890$; number of correct touched tiles: $BF_{01} = 3.236$).

\begin{table}[!t]
    \caption{The table shows the descriptive values of the task performance measures (\# refers to number).}
    \adjustbox{max width=\columnwidth}{
    \label{tab:descriptive-results-performance}
    \begin{tabular}{lcc}
    \toprule
    & \multicolumn{2}{c}{\ivr System} \\
    & Old & New \\
    \cmidrule{2-3}
    & $M$ ($SD$) & $M$ ($SD$) \\
    \midrule
    \textbf{Hallway Task} \\
    \hspace{4mm}\# of walls passed & 5.44 (2.97) & 5.27 (2.68) \\
    \hspace{4mm}\# of walls hit & 8.23 (2.15) & 8.37 (1.94) \\
    \textbf{Wire Task} \\
    \hspace{4mm} Wire distance (m) & 5.08 (1.99) & 5.16 (1.95) \\
    \hspace{4mm}\# of times wire left & 0.43 (0.53) & 0.49 (0.46) \\
    \textbf{Reaction Wall Task} \\
    \hspace{4mm}Reaction time (s) & 1.36 (0.35) & 1.31 (0.25) \\
    \hspace{4mm}\# of correct touched targets & 38.68 (8.67) & 39.03 (7.12) \\
    \textbf{Dance Task} \\
    \hspace{4mm}Reaction time (s) & 1.53 (0.51) & 1.46 (0.30) \\
    \hspace{4mm}\# of correct touched tiles & 20.45 (4.87) & 20.55 (2.81) \\

    \bottomrule
    \end{tabular}}
\end{table}

\begin{table}[htbp]
\caption{This table shows significant Post-hoc Wilcoxon signed-rank test results for task load comparisons between the different tasks. Statistical significance indicators: ${^{*}\,p<0.05}$; ${^{\dag}\,p< 0.01}$; ${^{\ddag}\,p< 0.001}$.}
\adjustbox{max width=\columnwidth}{
\label{tab:wilcoxon_taskload1}
\begin{tabular}{ll}
\toprule
Task Comparison & Test Statistics \\
\midrule
\textbf{Overall Task Load} \\
\hspace{4mm}Dance Task vs. Mirror Task & $W(40) = 820$, $p < .001^{\ddag}$ \\
\hspace{4mm}Dance Task vs. Reaction Wall Task & $W(40) = 593$, $p = .013^{*}$ \\
\hspace{4mm}Hallway Task vs. Mirror Task & $W(40) = 741$, $p < .001^{\ddag}$ \\
\hspace{4mm}Hallway Task vs. Wire Task & $W(40) = 168$, $p = .02^{*}$ \\
\hspace{4mm}Mirror Task vs. Reaction Wall Task & $W(40) = 24$, $p < .001^{\ddag}$ \\
\hspace{4mm}Mirror Task vs. Wire Task & $W(40) = 1$, $p < .001^{\ddag}$ \\
\hspace{4mm}Reaction Wall Task vs. Wire Task & $W(40) = 104$, $p < .001^{\ddag}$ \\    
\midrule

\textbf{Mental Demand} \\
\hspace{4mm}Dance Task vs. Hallway Task & $W(40) = 508$, $p = .003^{\dag}$ \\
\hspace{4mm}Dance Task vs. Mirror Task & $W(40) = 618$, $p < .001^{\ddag}$ \\
\hspace{4mm}Dance Task vs. Wire Task & $W(40) = 67.5$, $p = .001^{\dag}$ \\
\hspace{4mm}Hallway Task vs. Mirror Task & $W(40) = 372$, $p = .04^{*}$ \\
\hspace{4mm}Hallway Task vs. Wire Task & $W(40) = 38$, $p < .001^{\ddag}$ \\
\hspace{4mm}Mirror Task vs. Reaction Wall Task & $W(40) = 78$, $p = .005^{\dag}$ \\
\hspace{4mm}Mirror Task vs. Wire Task & $W(40) = 4$, $p < .001^{\ddag}$ \\
\hspace{4mm}Reaction Wall Task vs. Wire Task & $W(40) = 80.5$, $p < .001^{\ddag}$ \\
    \midrule

\textbf{Physical Demand} \\
\hspace{4mm}Dance Task vs. Mirror Task & $W(40) = 769$, $p < .001^{\ddag}$ \\
\hspace{4mm}Dance Task vs. Reaction Wall Task & $W(40) = 444$, $p = .001^{\dag}$ \\
\hspace{4mm}Dance Task vs. Wire Task & $W(40) = 659$, $p < .001^{\ddag}$ \\
\hspace{4mm}Hallway Task vs. Mirror Task & $W(40) = 589$, $p < .001^{\ddag}$ \\
\hspace{4mm}Hallway Task vs. Reaction Wall Task & $W(40) = 433$, $p = .016^{*}$ \\
\hspace{4mm}Hallway Task vs. Wire Task & $W(40) = 630$, $p < .001^{\ddag}$ \\
\hspace{4mm}Mirror Task vs. Reaction Wall Task & $W(40) = 92.5$, $p < .001^{\ddag}$ \\
    \midrule

\textbf{Frustration} \\
\hspace{4mm}Dance Task vs. Hallway Task & $W(40) = 534$, $p = .003^{\dag}$ \\
\hspace{4mm}Dance Task vs. Mirror Task & $W(40) = 478$, $p < .001^{\ddag}$ \\
\hspace{4mm}Dance Task vs. Reaction Wall Task & $W(40) = 394$, $p = .009^{\dag}$ \\
\hspace{4mm}Dance Task vs. Wire Task & $W(40) = 138$, $p = .023^{*}$ \\
\hspace{4mm}Hallway Task vs. Wire Task & $W(40) = 34$, $p < .001^{\ddag}$ \\
\hspace{4mm}Mirror Task vs. Wire Task & $W(40) = 30.5$, $p < .001^{\ddag}$ \\
\hspace{4mm}Reaction Wall Task vs. Wire Task & $W(40) = 34$, $p < .001^{\ddag}$ \\
    \midrule

\textbf{Temporal Demand} \\
\hspace{4mm}Dance Task vs. Mirror Task & $W(40) = 595$, $p < .001^{\ddag}$ \\
\hspace{4mm}Hallway Task vs. Mirror Task & $W(40) = 658$, $p < .001^{\ddag}$ \\
\hspace{4mm}Hallway Task vs. Wire Task & $W(40) = 616$, $p = .016^{*}$ \\
\hspace{4mm}Mirror Task vs. Reaction Wall Task & $W(40) = 16.5$, $p < .001^{\ddag}$ \\
\hspace{4mm}Mirror Task vs. Wire Task & $W(40) = 23.5$, $p < .001^{\ddag}$ \\
\hspace{4mm}Reaction Wall Task vs. Wire Task & $W(40) = 518$, $p = .037^{*}$ \\
    \midrule

\textbf{Performance} \\
\hspace{4mm}Dance Task vs. Mirror Task & $W(40) = 632$, $p < .001^{\ddag}$ \\
\hspace{4mm}Dance Task vs. Reaction Wall Task & $W(40) = 512$, $p = .002^{\dag}$ \\
\hspace{4mm}Dance Task vs. Wire Task & $W(40) = 95.5$, $p < .001^{\ddag}$ \\
\hspace{4mm}Hallway Task vs. Mirror Task & $W(40) = 544$, $p = .009^{\dag}$ \\
\hspace{4mm}Hallway Task vs. Wire Task & $W(40) = 49.5$, $p < .001^{\ddag}$ \\
\hspace{4mm}Mirror Task vs. Reaction Wall Task & $W(40) = 71$, $p = .005^{\dag}$ \\
\hspace{4mm}Mirror Task vs. Wire Task & $W(40) = 28.5$, $p < .001^{\ddag}$ \\
\hspace{4mm}Reaction Wall Task vs. Wire Task & $W(40) = 55$, $p < .001^{\ddag}$ \\
    \midrule

\textbf{Effort} \\
\hspace{4mm}Dance Task vs. Mirror Task & $W(40) = 699$, $p < .001^{\ddag}$ \\
\hspace{4mm}Hallway Task vs. Mirror Task & $W(40) = 733$, $p < .001^{\ddag}$ \\
\hspace{4mm}Mirror Task vs. Reaction Wall Task & $W(40) = 17$, $p < .001^{\ddag}$ \\
\hspace{4mm}Mirror Task vs. Wire Task & $W(40) = 9$, $p < .001^{\ddag}$ \\
\bottomrule
\end{tabular}}
\end{table}

\begin{table}[!t]
    \caption{Mann-Whitney U test results for task performance measures between old and new integrated VR systems (\# refers to number).}
    \adjustbox{max width=\columnwidth}{
    \label{tab:wilcoxon_tests}
    \begin{tabular}{lc}
    \toprule
    Task & Test Statistics \\
    \midrule
    \textbf{Hallway Task} \\
    \hspace{4mm}\# of walls passed & $U(19,20) = 195$, $p = .899$ \\
    \hspace{4mm}\# of walls hit & $U(19,20) = 181$, $p = .811$ \\
    \midrule

    \textbf{Wire Task} \\
    \hspace{4mm}Wire distance (m) & $U(18,19) = 159$, $p = .730$ \\
    \hspace{4mm}\# of times wire left  & $U(18,19) = 146.5$, $p = .450$ \\
    \midrule

    \textbf{Reaction Wall Task} \\
    \hspace{4mm}Reaction time (s) & $U(20,20) = 210$, $p = .799$ \\
    \hspace{4mm}\# of correct touched targets & $U(20,20) = 195$, $p = .903$ \\
    \midrule

    \textbf{Dance Task} \\
    \hspace{4mm}Reaction Time (s) & $U(20,20) = 204$, $p = .829$ \\
    \hspace{4mm}\# of correct touched tiles & $U(20,20) = 191.5$, $p = .920$ \\
    \bottomrule
    \end{tabular}}
\end{table}

\subsection{Qualitative Analysis}
We conducted a thematic analysis of the interviews, organizing participants' responses by question using sticky notes to identify recurring themes and patterns \cite{braun2006qualitativeevaluation}.
One focus of the interviews was participants' subjective assessment of the comfort of the \ivr system.
For the old \ivr system, 9 participants described it as uncomfortable, 3 were neutral, and 6 found it comfortable.
Discomfort was primarily attributed to the weight of the headset, as well as limitations in movement caused by the sensors and cable of the HMD, which participants found restrictive during tasks.
For the new \ivr system, 14 participants also described it as uncomfortable, 4 were neutral, and 14 found it comfortable.
Participants also noted issues with headset weight and limitations in movement due to the sensors and the cable of the HMD.
With both \ivr systems, in particular, the weight of the HMD was reported as a source of discomfort by 15 participants in total, specifically during the Dance Task, where participants had to repeatedly look down at their feet.

%% file: sections/05-discussion.tex
\section{Discussion}

Our study focuses on whether two different generations of \ivr systems affect user experience in VR. 
Therefore, we investigated whether technical improvements of \ivr systems (e.g., latency, tracking fidelity, frame rate, resolution, field of view), which result in higher immersion, also lead to an increased user experience and task performance. 
We adopted a user-centered approach, as technical improvements do not necessarily translate into better user experiences, and subjective evaluations remain essential to understand how system differences shape perception \cite{yun2023animationfidelityselfavatar}.
Our study employed a $2 \times 5$ mixed design.
Participants used either a state-of-the-art \ivr system or a nearly 10-year-old counterpart while performing five tasks in randomized order.
These tasks represent different use cases and target different body parts.
A key aspect was to evaluate the \ivr systems in their complete, commercially available as-is configurations. 
This choice enhances ecological validity and practical relevance, reflecting how most researchers and users encounter these systems without modifying individual components.
Simulator sickness was included as a control variable. No significant differences emerged in pre- and post-measures across systems, indicating that it did not act as a confound.

\subsection{Impact of \ivr System}
With our hypotheses, we expected that immersive \ivr system would result in a higher quality regarding perceived presence~\cite{cummings2016immersive, merz2024universalaccess}, sense of embodiment~\cite{born2019exergamesimmersionembodiment}, appearance and behavior plausibility~\cite{wolf2022plausbilitydisplay}, task load~\cite{tong2023towards}, and task performance~\cite{yun2023animationfidelityselfavatar}.
Overall, the new \ivr system delivers higher-resolution, higher-refresh-rate displays, broader tracking coverage, and lower motion-to-photon latency, collectively yielding higher objective immersion, which led us to expect differences in user experiences~\cite{cummings2016immersive, slater1997immersionandpresence}.
However, we reject all our hypotheses \textbf{H1, H2, H3, H4}, and \textbf{H5}.
Despite higher immersion, the new \ivr system did not improve the user experience qualia as expected, and our results did not show significant differences between the two systems.
Additionally, the qualitative feedback of the participants showed a similar perceived comfort between the two systems.

While prior work has compared different display types~\cite{cummings2016immersive}, there is a lack of work directly comparing different HMDs in terms of their influence on user experience.
It is possible that both \ivr systems were too similar in their technical properties of the display to elicit a difference in user experience.
System latency provides additional support that the \ivr systems were too similar.
While previous work showed that increased latency can negatively impact user experience~\cite{Caserman2020SurveyFullBody, Kasahara2017MalleableEmbodimentChanging, yun2023animationfidelityselfavatar}, Waltemate et al.~\cite{waltemate2016latency} reported that declines in the sense of embodiment begin to occur only at latencies exceeding \unitmillisecond{125}.
The new \ivr system exhibits substantially lower latency (M = \unitmillisecond{16.15}) than the old system (M = \unitmillisecond{85.94}).
However, despite its higher latency, the descriptive values support that the old \ivr system remained in an acceptable range and effectively \enquote{good enough} to support a comparable user experience.

Furthermore, we expected the additional and improved sensors used for tracking in the new \ivr system to enhance the user experience.
Unlike the old system, which relies on two base stations, the new system uses four base stations with a broader field of view and a higher refresh rate.
This configuration provides a comprehensive coverage of the tracking space and should result in more accurate tracking of HMD and trackers.
Additionally, the new system is theoretically less susceptible to occlusion, which can compromise tracking fidelity.
However, occlusion probably did not substentially impact our study.
The lab environment was free from obstacles and the tasks did not require participants to adopt postures that would occlude the trackers from all base stations with their physical bodies.
Thus, the lower number of base stations in the old system may not have negatively affected tracking fidelity in our specific setup.

Overall, ANOVA analyses do not allow the interpretation that no significant differences translate to no effect, particularly given our small sample size. 
However, taken together, our statistical analyses consistently indicate that any potential effect is likely negligible: (1) Bayesian ANOVAs provided evidence that the null hypothesis is more likely than our model of the \ivr system across all measures, (2) trimmed-means ANOVAs showed low effect sizes for the \ivr systems, suggesting that any undetected effect would be very small.
As a result, we cannot confirm our hypotheses that different generations of \ivr systems impact user experience or task performance.
This suggests that technical improvements, while objectively determinable, may not translate directly into improved user experience or performance.
Supporting this interpretation, participants reported comparable comfort across systems.
Several participants mentioned the HMD's weight and the restricted movements from cables and tracker placement as limiting factors for both systems.
Thus, despite technological advances, the perceptual benefits of the tested newer \ivr system appear to be minimal. 
Our qualitative results show that future improvements should consider wearability and reducing physical demand, as comfort may be more critical than incremental technical upgrades.
To sum up, our findings suggest that while there are technical improvements over generations of \ivr systems, their perceptual impact appears to be limited and additional focus must be placed on further factors such as comfort.

\subsection{Impact of Task}
We did not state hypotheses that the user experience is different for the tasks.
However, we performed an exploratory evaluation to determine whether there is a difference between tasks or an interaction effect between tasks and \ivr systems.
Task load ratings differed significantly between tasks, which is expected, since each task represented a different use case and required different user actions.
Interestingly, when comparing our tasks, we found a significant main effect for spatial presence,  general presence, and all subscales of sense of embodiment. This can be explained by differing task requirements regarding the user's focus. For example, during the reaction wall or dance tasks, participants interacted with only part of their body, whereas in the hallway task, they had to move their entire avatar to complete the task, which appeared to result in a higher sense of embodiment. Furthermore, in the dance task, participants had to tilt their heads downward to look at their feet. This posture made the weight of the HMD more noticeable, as mentioned in the qualitative feedback.
Compared to the other tasks, in the mirror task, the whole body is visible through the mirror, becomes the main focus, and is observed more holistically for a longer period. This may direct more attention to the avatar’s appearance and behavior, potentially increasing critical avatar evaluation. This suggests that inaccurate tracking or pose mismatches may be more noticeable in the mirror task, potentially diminishing virtual body ownership in particular.
This contrasts with the other, more goal-oriented tasks, where attention is directed toward the environment and task performance. In these tasks, the virtual body primarily serves as a means to accomplish the task, and only parts of the body are in the field of view. Additionally, prior work showed that avatars with facial animations are rated with a higher sense of embodiment~\cite{kullmann2025facialexpressioneffect}. In our study, the avatar’s non-animated face is only visible in the mirror task, which may have further diminished the experienced sense of embodiment in the mirror task.

Overall, while previous work often used mirror exposure to induce the feeling of having and controlling a body~\cite{inoue2021mirror, roth2018avatar, waltemate2018impact, wolf2021embodiment}, our results suggest that other, more goal-oriented tasks can elicit a higher sense of embodiment. Therefore, in certain use cases where the avatar's appearance is not central, it can be beneficial to choose other tasks to induce the feeling of having and controlling a body.

\subsection{Implications}
Our results have several important implications for research and practice using \ivr systems. 
Contrary to our hypotheses and previous literature suggesting a positive relationship between increased immersion and user experience or task performance~\cite{born2019exergamesimmersionembodiment,cummings2016immersive,merz2024universalaccess,tong2023towards,wolf2022plausbilitydisplay,yun2023animationfidelityselfavatar}, we found no significant differences between the old and new \ivr systems, despite technological advancements.
This absence of effects supports the complex and potentially non-linear relationship between technical improvements and user perception~\cite{latoschik2019alone,Caserman2020SurveyFullBody, Kimmel2024KineticConnectionsExploring}.  
This suggests that improvements in technological parameters, such as latency, tracking fidelity, frame rate, resolution, and field of view alone do not guarantee improved subjective experiences, aligning with recent findings that technical performance metrics do not consistently translate into perceived user benefits~\cite{yun2023animationfidelityselfavatar}.
At the same time, our results indicate that the findings derived from older \ivr systems still hold relevance, reinforcing the applicability of prior research.
We therefore expect that repeating prior studies with a state-of-the-art \ivr system would yield comparable results regarding the user experience.

Furthermore, these findings emphasize the need for user-centered evaluation when evaluating \ivr systems. 
Factors such as system familiarity, comfort, user expectations, and demographic factors (e.g., age or previous VR experience) might influence user experience and task performance more strongly than incremental technical upgrades.
For example, VR experts who work with different \ivr systems on a daily basis might be more sensitive to the lower quality of older systems.
Furthermore, prior work showed age-related sensory and perceptual effects in VR~\cite{dilanchian2021pilot}.
Taken together, these findings suggest that immediate hardware upgrades may not be necessary for practical use, and decisions should consider factors beyond technical specifications.

Our findings also highlight the role of task design in shaping user experience in VR. 
While mirror exposure is commonly used to enhance the sense of embodiment~\cite{inoue2021mirror}, our results suggest that tasks requiring active and full-body engagement may elicit a stronger sense of embodiment. 
This is interesting in scenarios where a sense of embodiment should be created, but observing the appearance of the avatar in a mirror is not necessarily important~\cite{born2019exergamesimmersionembodiment}.
In such cases, movement-focused tasks may be more effective in strengthening the user's sense of having and controlling a virtual body.

Overall, these implications underline the complexity of user experience in VR.
Advancing \ivr technologies requires a balanced approach that integrates technological progress with user-centered design and considers contextual factors such as demographics and task-specific demands, influencing perception and interaction.

\subsection{Limitations and Future Work}
First, our study was conducted with a homogeneous sample consisting predominantly of university students of the same ethnicity. Hence, the generalization of our findings to broader and more diverse populations is limited. Future research should seek to replicate and validate these results using a more representative sample.

Second, we manipulated the entire \ivr system, including HMD, providing both display and head pose tracking, and additional sensor capabilities in their complete commercially available as-is configuration.
Although this ensures high ecological validity, it prevents isolating the influence of specific technical factors on user experience and task performance.
While our work did not aim to disentangle the contributions of individual immersive components, the overall improvements in the new \ivr system did not lead to measurable improvements in user experience. 
This aligns with previous work that indicates that perceptual thresholds may buffer small technical changes~\cite{waltemate2016latency}. 
Future research should therefore conduct component-level analyses to disentangle these contributions and determine which system parameters most strongly shape user experience and task performance.
These studies could build on work such as from Nakano et al.~\cite{Nakano2021HeadMountedDisplay, Nakano2025AvatarsShouldWe} and Brübach et al.~\cite{Bruebach2024InfluenceLowResolution}, who demonstrated how targeted manipulations of the display characteristics (e.g., vertical field of view) can influence sense of embodiment and presence, whereas others may not. 
Additionally, keeping the HMD of the \ivr system constant while manipulating additional sensor capabilities only influences the user experience~\cite{yun2023animationfidelityselfavatar}. 
However, there is still a gap in singling out these additional sensor capabilities. 
Disentangling them systematically would help clarify which technological upgrades impact user experience.

Finally, we could not fully replicate the software configuration of earlier research involving the older \ivr system, due to well-known replication obstacles~\cite{grubel:2023a}.
The Vive hardware receives software updates over time that may affect the way position and rotation are calculated.
We used the same software stack for both systems, including the same version of SteamVR, the developed application, and the IK framework.
Improvements through software updates could have contributed to higher immersion, user experience, and performance in the older \ivr system than what would have been observed with older software versions.

%% file: sections/06-conclusion.tex
\section{Conclusion}
Our study compared two generations of \ivr systems to test whether technological advances (e.g., latency, tracking fidelity, frame rate, resolution, field of view) enhance user experience and task performance. 
We adopted a user-centered study design, recognizing that technical specifications do not necessarily translate into better user experiences, and evaluated both systems in their complete, commercially available as-is configurations to ensure ecological validity and practical relevance. 

Across presence, sense of embodiment, appearance and behavior plausibility, task load, and task performance, we found no significant differences, with Bayesian analyses supporting the null hypothesis.
The observed effect sizes were minimal, indicating negligible practical differences between the two systems regarding user experience.
These findings suggest that despite technological advances, user experience did not differ significantly between the two generations of \ivr systems.
Therefore, conclusions from previous studies employing our old \ivr system or systems with comparable immersive properties still hold relevance for our new \ivr system or systems with comparable immersive properties, although minor improvements in system usability and comfort might exist. 

However, we believe that our findings cannot easily be generalized to \ivr systems with different immersive properties, as the relationship between system properties and user experience can be quite complex.
This is affected by the magnitude of technological differences and advancements (e.g., sensitivity, accuracy, precision, resolution, range, latency, adaptability, etc.).
Our data further indicated that user-centered factors, such as wearability, familiarity, and task design, may influence user experience more strongly than incremental technical upgrades. 
Future work should disentangle the contributions of individual system components and prioritize user-centered aspects in shaping user experience in \ivr.

%% file: references.bib
@Article{ourbodytrackingcomparison,
  author    = {Tschanter, Jonathan and Merz, Christian and Fiedler, Marie Luisa and Wienrich, Carolin and Latoschik, Marc Erich},
  title     = {Use Case Matters: {Comparing} the User Experience and Task Performance Across Tasks for Embodied Interaction in {VR}},
  journal   = {IEEE Transactions on Visualization and Computer Graphics},
  year      = {2026},
  note = {Conditionally accepted}
}

@Article{Badler1993RealTimeControl,
  author    = {Badler, Norman I. and Hollick, Michael J. and Granieri, John P.},
  journal   = {Presence: Teleoperators and Virtual Environments},
  title     = {Real-Time Control of a Virtual Human Using Minimal Sensors},
  year      = {1993},
  issn      = {1054-7460},
  number    = {1},
  pages     = {82--86},
  volume    = {2},
  doi       = {10.1162/pres.1993.2.1.82},
  publisher = {MIT Press - Journals},
}

@Article{Nakano2025AvatarsShouldWe,
  author    = {Nakano, Kizashi and Narumi, Takuji},
  journal   = {IEEE Transactions on Visualization and Computer Graphics},
  title     = {Avatars, Should We Look at Them Directly or Through a Mirror?: Effects of Avatar Display Method on Sense of Embodiment and Gaze},
  year      = {2025},
  issn      = {2160-9306},
  number    = {5},
  pages     = {2912--2922},
  volume    = {31},
  doi       = {10.1109/tvcg.2025.3549545},
  publisher = {IEEE},
}

@Article{Nakano2021HeadMountedDisplay,
  author    = {Nakano, Kizashi and Isoyama, Naoya and Monteiro, Diego and Sakata, Nobuchika and Kiyokawa, Kiyoshi and Narumi, Takuji},
  journal   = {IEEE Transactions on Visualization and Computer Graphics},
  title     = {Head-Mounted Display with Increased Downward Field of View Improves Presence and Sense of Self-Location},
  year      = {2021},
  issn      = {2160-9306},
  number    = {11},
  pages     = {4204--4214},
  volume    = {27},
  doi       = {10.1109/tvcg.2021.3106513},
  publisher = {IEEE},
}

@InProceedings{Bruebach2024InfluenceLowResolution,
  author     = {Brübach, Larissa and Röhm, Marius and Westermeier, Franziska and Wienrich, Carolin and Latoschik, Marc Erich},
  booktitle  = {30th ACM Symposium on Virtual Reality Software and Technology},
  title      = {The Influence of a Low-Resolution Peripheral Display Extension on the Perceived Plausibility and Presence},
  year       = {2024},
  pages      = {1--10},
  publisher  = {ACM},
  series     = {VRST ’24},
  collection = {VRST ’24},
  doi        = {10.1145/3641825.3687713},
}

@InProceedings{Bartl2022EffectsAvatarEnvironment,
  author    = {Bartl, Andrea and Merz, Christian and Roth, Daniel and Latoschik, Marc Erich},
  booktitle = {2022 ISMAR},
  title     = {The Effects of Avatar and Environment Design on Embodiment, Presence, Activation, and Task Load in a Virtual Reality Exercise Application},
  year      = {2022},
  pages     = {260--269},
  publisher = {IEEE},
  doi       = {10.1109/ismar55827.2022.00041},
}

@InProceedings{Boulic1997Integrationmotioncontrol,
  author     = {Boulic, Ronan and Bécheiraz, Pascal and Emering, Luc and Thalmann, Daniel},
  booktitle  = {4th ACM Symposium on Virtual Reality Software and Technology},
  title      = {Integration of motion control techniques for virtual human and avatar real-time animation},
  year       = {1997},
  publisher  = {ACM},
  series     = {VRST ’97},
  collection = {VRST ’97},
  doi        = {10.1145/261135.261156},
}

@Article{Gutierrez2024ComparingOpticalCustom,
  author    = {Gutierrez, Manuel and Gomez, Britam and Retamal, Gustavo and Peña, Guisella and Germany, Enrique and Ortega-Bastidas, Paulina and Aqueveque, Pablo},
  journal   = {Technologies},
  title     = {Comparing Optical and Custom {IoT} Inertial Motion Capture Systems for Manual Material Handling Risk Assessment Using the {NIOSH} Lifting Index},
  year      = {2024},
  issn      = {2227-7080},
  number    = {10},
  pages     = {180},
  volume    = {12},
  doi       = {10.3390/technologies12100180},
  publisher = {MDPI AG},
}

@InProceedings{Holzwarth2021ComparingAccuracyPrecision,
  author     = {Holzwarth, Valentin and Gisler, Joy and Hirt, Christian and Kunz, Andreas},
  booktitle  = {5th International Conference on Virtual and Augmented Reality Simulations},
  title      = {Comparing the Accuracy and Precision of {SteamVR} Tracking 2.0 and {Oculus Quest} 2 in a Room Scale Setup},
  year       = {2021},
  publisher  = {ACM},
  series     = {ICVARS ’21},
  collection = {ICVARS ’21},
  doi        = {10.1145/3463914.3463921},
}

@Article{Kimmel2024KineticConnectionsExploring,
  author    = {Kimmel, Simon and Landwehr, Eric and Heuten, Wilko},
  journal   = {Proceedings of the ACM on Human-Computer Interaction},
  title     = {Kinetic Connections: {E}xploring the Impact of Realistic Body Movements on Social Presence in Collaborative Virtual Reality},
  year      = {2024},
  issn      = {2573-0142},
  number    = {CSCW2},
  pages     = {1--30},
  volume    = {8},
  doi       = {10.1145/3686910},
  publisher = {ACM},
}

@Article{Merker2023MeasurementAccuracyHTC,
  author    = {Merker, Sebastian and Pastel, Stefan and Bürger, Dan and Schwadtke, Alexander and Witte, Kerstin},
  journal   = {Sensors},
  title     = {Measurement Accuracy of the {HTC VIVE} Tracker 3.0 Compared to {V}icon System for Generating Valid Positional Feedback in Virtual Reality},
  year      = {2023},
  issn      = {1424-8220},
  number    = {17},
  pages     = {7371},
  volume    = {23},
  doi       = {10.3390/s23177371},
  publisher = {MDPI AG},
}

@InProceedings{Mohler2008fullbodyavatar,
  author     = {Mohler, Betty J. and Bülthoff, Heinrich H. and Thompson, William B. and Creem-Regehr, Sarah H.},
  booktitle  = {5th Symposium on Applied Perception in Graphics and Visualization},
  title      = {A full-body avatar improves egocentric distance judgments in an immersive virtual environment},
  year       = {2008},
  publisher  = {ACM},
  series     = {APGV ’08},
  collection = {APGV ’08},
  doi        = {10.1145/1394281.1394323},
}

@Article{Peck2013Puttingyourselfskin,
  author    = {Peck, Tabitha C. and Seinfeld, Sofia and Aglioti, Salvatore M. and Slater, Mel},
  journal   = {Consciousness and Cognition},
  title     = {Putting yourself in the skin of a black avatar reduces implicit racial bias},
  year      = {2013},
  issn      = {1053-8100},
  number    = {3},
  pages     = {779--787},
  volume    = {22},
  doi       = {10.1016/j.concog.2013.04.016},
  publisher = {Elsevier BV},
}

@Article{Caserman2020SurveyFullBody,
  author    = {Caserman, Polona and Garcia-Agundez, Augusto and Göbel, Stefan},
  journal   = {IEEE Transactions on Visualization and Computer Graphics},
  title     = {A Survey of Full-Body Motion Reconstruction in Immersive Virtual Reality Applications},
  year      = {2020},
  issn      = {2160-9306},
  number    = {10},
  pages     = {3089--3108},
  volume    = {26},
  doi       = {10.1109/tvcg.2019.2912607},
  publisher = {IEEE},
}

@InProceedings{Young2015Dyadicinteractionsavatars,
  author     = {Young, Mary K. and Rieser, John J. and Bodenheimer, Bobby},
  booktitle  = {Proceedings of the ACM SIGGRAPH Symposium on Applied Perception},
  title      = {Dyadic interactions with avatars in immersive virtual environments: high fiving},
  year       = {2015},
  pages      = {119--126},
  publisher  = {ACM},
  series     = {SAP ’15},
  collection = {SAP ’15},
  doi        = {10.1145/2804408.2804410},
}

@InProceedings{Hepke2024DevelopmentValidation3D,
  author     = {Hepke, Nikolai and Scherer, Moritz and Lohscheller, Jörg and Müller, Steffen and Weyers, Benjamin},
  booktitle  = {30th ACM Symposium on Virtual Reality Software and Technology},
  title      = {Development and Validation of a {3D} Pose Tracking System towards {XR} Home Training to Relieve Back Pain},
  year       = {2024},
  pages      = {1--11},
  publisher  = {ACM},
  series     = {VRST ’24},
  collection = {VRST ’24},
  doi        = {10.1145/3641825.3687746},
}

@InProceedings{Kasahara2017MalleableEmbodimentChanging,
  author     = {Kasahara, Shunichi and Konno, Keina and Owaki, Richi and Nishi, Tsubasa and Takeshita, Akiko and Ito, Takayuki and Kasuga, Shoko and Ushiba, Junichi},
  booktitle  = {2017 Conference on Human Factors in Computing Systems},
  title      = {Malleable Embodiment: Changing Sense of Embodiment by Spatial-Temporal Deformation of Virtual Human Body},
  year       = {2017},
  publisher  = {ACM},
  series     = {CHI ’17},
  collection = {CHI ’17},
  doi        = {10.1145/3025453.3025962},
}

@InProceedings{Lugrin2013Usabilitybenchmarksmotion,
  author     = {Lugrin, Jean-Luc and Wiebusch, Dennis and Latoschik, Marc Erich and Strehler, Alexander},
  booktitle  = {19th ACM Symposium on Virtual Reality Software and Technology},
  title      = {Usability benchmarks for motion tracking systems},
  year       = {2013},
  pages      = {49--58},
  publisher  = {ACM},
  series     = {VRST ’13},
  collection = {VRST ’13},
  doi        = {10.1145/2503713.2503730},
}

@article{dilanchian2021pilot,
  title={A pilot study exploring age differences in presence, workload, and cybersickness in the experience of immersive virtual reality environments},
  author={Dilanchian, Andrew T and Andringa, Ronald and Boot, Walter R},
  journal={Frontiers in Virtual Reality},
  volume={2},
  pages={736793},
  year={2021},
  doi= {https://doi.org/10.3389/frvir.2021.736793},
  publisher={Frontiers Media SA}
}

@INPROCEEDINGS{merz2024voice,
  author={Merz, Christian and Wienrich, Carolin and Latoschik, Marc Erich},
  booktitle={2024 IEEE International Symposium on Mixed and Augmented Reality (ISMAR)}, 
  title={Does Voice Matter? {T}he Effect of Verbal Communication and Asymmetry on the Experience of Collaborative Social XR}, 
  year={2024},
  pages={1127-1136},
  doi={10.1109/ISMAR62088.2024.00129}}

@book{cohen1988statistical,
  title     = {Statistical Power Analysis for the Behavioral Sciences},
  author    = {Cohen, Jacob},
  year      = {1988},
  publisher = {Routledge},
  address   = {Hillsdale, NJ},
  doi       = {10.4324/9780203771587},
}

@ARTICLE{kullmann2025facialexpressioneffect,
  author={Kullmann, Peter and Schell, Theresa and Menzel, Timo and Botsch, Mario and Latoschik, Marc Erich},
  journal={IEEE Transactions on Visualization and Computer Graphics}, 
  title={Coverage of Facial Expressions and Its Effects on Avatar Embodiment, Self-Identification, and Uncanniness}, 
  year={2025},
  volume={31},
  number={5},
  pages={3613-3622},
  doi={10.1109/TVCG.2025.3549887}}

@INPROCEEDINGS {tong2023towards,
author = {W. Tong and M. Xia and K. Wong and D. A. Bowman and T. Pong and H. Qu and Y. Yang},
booktitle = {2023 IEEE VR},
title = {Towards an Understanding of Distributed Asymmetric Collaborative Visualization on Problem-solving},
year = {2023},
volume = {},
issn = {},
pages = {387-397},
doi = {10.1109/VR55154.2023.00054},
publisher = {IEEE},
address = {Los Alamitos, CA, USA},
}

@Article{cummings2016immersive,
  author     = {Cummings, James J. and Bailenson, Jeremy N.},
  journal    = {Media Psychology},
  title      = {How Immersive Is Enough? {A} Meta-Analysis of the Effect of Immersive Technology on User Presence},
  year       = {2015},
  issn       = {1532-785X},
  number     = {2},
  pages      = {272--309},
  volume     = {19},
  doi        = {10.1080/15213269.2015.1015740},
  publisher  = {Informa UK Limited},
}

@inproceedings{kern2023reality,
  title={Reality stack i/o: {A} versatile and modular framework for simplifying and unifying {XR} applications and research},
  author={Kern, Florian and Latoschik, Marc Erich},
  booktitle={2023 IEEE International Symposium on Mixed and Augmented Reality Adjunct (ISMAR-Adjunct)},
  pages={74--76},
  year={2023},
  publisher={IEEE},
  doi={10.1109/ISMAR-Adjunct60411.2023.00023}
}

@inproceedings{merz2024pipelining,
  title={Pipelining Processors for Decomposing Character Animation},
  author={Merz, Christian and Tschanter, Jonathan and Kern, Florian and Lugrin, Jean-Luc and Wienrich, Carolin and Latoschik, Marc Erich},
  booktitle  = {30th ACM Symposium on Virtual Reality Software and Technology},
  pages={1--2},
  publisher  = {ACM},
  series     = {VRST ’24},
  collection = {VRST ’24},
  year={2024},
  doi={10.1145/3641825.3689533}
}

@article{do2023valid,
AUTHOR={Do, Tiffany D. and Zelenty, Steve and Gonzalez-Franco, Mar and McMahan, Ryan P.},   
TITLE={VALID: a perceptually validated Virtual Avatar Library for Inclusion and Diversity},      
JOURNAL={Frontiers in Virtual Reality},      
VOLUME={4},           
YEAR={2023},       
URL={https://www.frontiersin.org/articles/10.3389/frvir.2023.1248915},       
DOI={10.3389/frvir.2023.1248915},      
ISSN={2673-4192}
}

@article{juan2009comparison,
  title={Comparison of the levels of presence and anxiety in an acrophobic environment viewed via {HMD} or CAVE},
  author={Juan, M Carmen and P{\'e}rez, David},
  journal={Presence: Teleoperators and virtual environments},
  volume={18},
  number={3},
  pages={232--248},
  year={2009},
  publisher={MIT Press One Rogers Street, Cambridge, MA 02142-1209, USA},
  doi={https://doi.org/10.1162/pres.18.3.232}
}

@article{slater2003note,
  title={A note on presence terminology},
  author={Slater, Mel},
  journal={Presence connect},
  volume={3},
  number={3},
  pages={1--5},
  year={2003}
}

@InProceedings{dollinger2019vitras,
  author    = {Döllinger, Nina and Wienrich, Carolin and Wolf, Erik and Latoschik, Marc Erich},
  booktitle = {Proc.\ MuC},
  title     = {{ViTraS} -- {V}irtual reality therapy by stimulation of modulated body image -- project outline},
  year      = {2019},
  address   = {Bonn},
  pages     = {1--6},
  publisher = {Gesellschaft für Informatik e.V.},
  doi       = {10.18420/muc2019-ws-633},
}

@Article{kilteni2012embodimentinvr,
  author  = {Kilteni, Konstantina and Groten, Raphaela and Slater, Mel},
  journal = {Presence: Teleoperators \& Virtual Environments},
  title   = {The sense of embodiment in virtual reality},
  year    = {2012},
  issn    = {1054-7460},
  number  = {4},
  pages   = {373-387},
  volume  = {21},
  doi     = {10.1162/PRES_a_00124 },
}

@techreport{witmer1994measuring,
  title={Measuring presence in virtual environments},
  author={Witmer, Bob G and Singer, Michael F},
  year={1994},
  institution={Army Research Inst for the Behavioral and Social Sciences Alexandria VA}
}

@inproceedings{mal2022virtual, 
      author={Mal, David and Wolf, Erik and Döllinger, Nina and Botsch, Mario and Wienrich, Carolin and Latoschik, Marc Erich},
  booktitle = {Proc.\ VR (VRW)},
      title={Virtual Human Coherence and Plausibility -- {T}owards a Validated Scale}, 
      year={2022},
      volume={},
      number={},
      pages={788-789},
      publisher = {IEEE},
      address = {New York, NY, USA},
      doi={10.1109/VRW55335.2022.00245  }
}

@inproceedings{mal2024vhpvr,
    author = {David Mal and Erik Wolf and Nina Döllinger and Mario Botsch and Carolin Wienrich and Marc Erich Latoschik},
    url = {https://downloads.hci.informatik.uni-wuerzburg.de/2024-chi-vhp-in-vr-preprint.pdf},
    year = {2024},
    booktitle = {CHI 24 Conference on Human Factors in Computing Systems Extended Abstracts},
    pages = {1-8},
    doi = {10.1145/3613905.3650773},
    title = {From 2{D}-Screens to {VR}: Exploring the Effect of Immersion on the Plausibility of Virtual Humans}
}

@article{latoschik2019alone,
    author = {Marc Erich Latoschik and Florian Kern and Jan-Philipp Stauffert and Andrea Bartl and Mario Botsch and Jean-Luc Lugrin},
    journal = {IEEE Transactions on Visualization and Computer Graphics (TVCG)},
    number = {5},
    url = {https://ieeexplore.ieee.org/document/8643417},
    year = {2019},
    pages = {2134-2144},
    volume = {25},
    doi = {10.1109/TVCG.2019.2899250},
    title = {Not Alone Here?! {S}calability and User Experience of Embodied Ambient Crowds in Distributed Social Virtual Reality}
}

@article{grubel:2023a,
	author = {Gr{\"u}bel, Jascha},
	doi = {10.3389/frvir.2023.1069423},
	issn = {2673-4192},
	journal = {Frontiers in Virtual Reality},
	title = {The design, experiment, analyse, and reproduce principle for experimentation in virtual reality},
	url = {https://www.frontiersin.org/journals/virtual-reality/articles/10.3389/frvir.2023.1069423},
	volume = {Volume 4 - 2023},
	year = {2023}}

@Article{braun2006qualitativeevaluation,
  author    = {Virginia Braun and Victoria Clarke},
  journal   = {Qualitative Research in Psychology},
  title     = {Using thematic analysis in psychology},
  year      = {2006},
  number    = {2},
  pages     = {77-101},
  volume    = {3},
  doi       = {10.1191/1478088706qp063oa},
  publisher = {Routledge},
  url       = {https://www.tandfonline.com/doi/abs/10.1191/1478088706qp063oa},
}

@inproceedings{roth2018avatar,
   author = {Roth, Daniel and Wienrich, Carolin},
   title = {Effects of Media Immersiveness on the Perception of Virtual Characters},
   booktitle = {15th Workshop on Virtual and Augmented Reality of the GI Special Interest Group VR/AR},
   year = {2018}
}

@article{roth2020veq,
   author = {Roth, Daniel and Latoschik, Marc Erich},
   title = {Construction of the virtual embodiment questionnaire {(VEQ)}},
   journal = {IEEE Transactions on Visualization and Computer Graphics},
   publisher = {IEEE},
   address = {New York, NY, USA},
   volume = {26},
   number = {12},
   pages = {3546-3556},
   ISSN = {1941-0506},
   DOI = {10.1109/TVCG.2020.3023603},
   year = {2020}
}

@Article{schubert2001experience,
  author    = {Schubert, Thomas and Friedmann, Frank and Regenbrecht, Holger},
  journal   = {Presence: Teleoperators \& Virtual Environments},
  title     = {The experience of presence: {F}actor analytic insights},
  year      = {2001},
  number    = {3},
  pages     = {266--281},
  volume    = {10},
  doi       = {10.1162/105474601300343603},
  publisher = {MIT Press},
}

@article{slater1997immersionandpresence,
	author = {Slater, Mel and Wilbur, Sylvia},
	title = {A framework for immersive virtual environments ({FIVE}): {S}peculations on the role of presence in virtual environments},
	journal = {Presence: Teleoperators \& Virtual Environments},
	volume = {6},
	number = {6},
	pages = {603-616},
	ISSN = {1054-7460},
	year = {1997},
	doi = {10.1162/pres.1997.6.6.603}
}

@ARTICLE{stauffert2020latencyreview,
AUTHOR={Stauffert, Jan-Philipp and Niebling, Florian and Latoschik, Marc Erich},   
TITLE={Latency and cybersickness: {I}mpact, causes, and measures. {A} review},      
JOURNAL={Frontiers in Virtual Reality},      
VOLUME={1},           
YEAR={2020},      
URL={https://www.frontiersin.org/articles/10.3389/frvir.2020.582204},       
DOI={10.3389/frvir.2020.582204},
}

@misc{morey2018bayesfactor,
	author = {Morey, Richard D. and Rouder, J. N.},
	year = {2018},
	title = {{BayesFactor: C}omputation of bayes factors for common designs},
	howpublished = {R package version 0.9.12-4.2},
}

@InProceedings{waltemate2016latency,
  author    = {Waltemate, Thomas and Senna, Irene and Hülsmann, Felix and Rohde, Marieke and Kopp, Stefan and Ernst, Marc and Botsch, Mario},
  booktitle={22th ACM Symposium on Virtual Reality Software and Technology},
  title     = {The impact of latency on perceptual judgments and motor performance in closed-loop interaction in virtual reality},
  year      = {2016},
  pages     = {27–35},
  publisher = {ACM},
  series     = {VRST ’16},
  collection = {VRST ’16},
  doi       = {10.1145/2993369.2993381},
  isbn      = {1450344917},
  url       = {https://doi.org/10.1145/2993369.2993381},
}

@Article{waltemate2018impact,
  author    = {Waltemate, Thomas and Gall, Dominik and Roth, Daniel and Botsch, Mario and Latoschik, Marc Erich},
  journal   = {IEEE Transactions on Visualization and Computer Graphics},
  title     = {The impact of avatar personalization and immersion on virtual body ownership, presence, and emotional response},
  year      = {2018},
  number    = {4},
  pages     = {1643--1652},
  volume    = {24},
  doi       = {10.1109/TVCG.2018.2794629},
  publisher = {IEEE},
  address = {New York, NY, USA},
}

@InProceedings{wolf2021embodiment,
  author    = {Wolf, Erik and Merdan, Nathalie and Döllinger, Nina and Mal, David and Wienrich, Carolin and Botsch, Mario and Latoschik, Marc Erich},
  booktitle = {2021 IEEE VR},
  title     = {The embodiment of photorealistic avatars influences female body weight perception in virtual reality},
  year      = {2021},
  pages     = {65-74},
  doi       = {10.1109/VR50410.2021.00027},
  publisher = {IEEE},
}

@inproceedings{wolf2022plausbilitydisplay, 
	author = {Wolf, Erik and Mal, David and Frohnapfel, Viktor and Döllinger, Nina and Wenninger, Stephan and Botsch, Mario and Latoschik, Marc Erich and Wienrich, Carolin},
	title = {Plausibility and perception of personalized virtual humans between virtual and augmented reality},
	year = {2022},
	booktitle = {2022 IEEE International Symposium on Mixed and Augmented Reality (ISMAR)},
	publisher = {IEEE},
    pages        = {489-498},
    doi          = {10.1109/ISMAR55827.2022.00065}
}

@article{inoue2021mirror,
	title={Virtual mirror and beyond: {T}he psychological basis for avatar embodiment via a mirror},
	author={Inoue, Yasuyuki and Kitazaki, Michiteru},
	journal={Journal of Robotics and Mechatronics},
	volume={33},
	number={5},
	pages={1004--1012},
	year={2021},
	doi={10.20965/jrm.2021.p1004}
}

@article{vox2021trackingcomparison,
    AUTHOR = {Vox, Jan P. and Weber, Anika and Wolf, Karen Insa and Izdebski, Krzysztof and Schüler, Thomas and König, Peter and Wallhoff, Frank and Friemert, Daniel},
    TITLE = {An Evaluation of Motion Trackers with Virtual Reality Sensor Technology in Comparison to a Marker-Based Motion Capture System Based on Joint Angles for Ergonomic Risk Assessment},
    JOURNAL = {Sensors},
    VOLUME = {21},
    YEAR = {2021},
    NUMBER = {9},
    ARTICLE-NUMBER = {3145},
    URL = {https://www.mdpi.com/1424-8220/21/9/3145},
    ISSN = {1424-8220},
    DOI = {10.3390/s21093145}
}

@article{mair2020wrs2,
  title={Robust statistical methods in {R using the WRS2} package},
  author={Mair, Patrick and Wilcox, Rand},
  journal={Behavior research methods},
  volume={52},
  number={2},
  year={2020},
  doi={10.3758/s13428-019-01246-w},
  pages =  {464--488},
}

@inproceedings{fiedler2023embodiment,
  title={Embodiment and personalization for self-identification with virtual humans},
  author={Fiedler, Marie Luisa and Wolf, Erik and D{\"o}llinger, Nina and Botsch, Mario and Latoschik, Marc Erich and Wienrich, Carolin},
  booktitle = {Proc.\ VR (VRW)},
  pages={799--800},
  year={2023},
  publisher = {IEEE},
  doi  = {10.1109/VRW58643.2023.00242}
}

@article {portingale2024review,
author={Portingale, Jade
and Krug, Isabel
and Liu, Hermione
and Kiropoulos, Litza
and Butler, David},
title={Your body, my experience: {A} systematic review of embodiment illusions as a function of and method to improve body image disturbance},
journal={Clinical Psychology: Science and Practice},
year={2024},
publisher={Educational Publishing Foundation},
address={US},
pages={1--15},
doi={10.1037/cps0000223},
url={https://doi.org/10.1037/cps0000223}
}

@ARTICLE{caserman2022impactfullbodytrainingpolice,
  author={Caserman, Polona and Schmidt, Philip and Göbel, Thorsten and Zinnäcker, Jonas and Kecke, André and Göbel, Stefan},
  journal={IEEE Transactions on Games}, 
  title={Impact of Full-Body Avatars in Immersive Multiplayer Virtual Reality Training for Police Forces}, 
  year={2022},
  volume={14},
  number={4},
  pages={706-714},
  doi={10.1109/TG.2022.3148791}
}

@inproceedings{smith2018communicationbehaviourvr,
author = {Smith, Harrison Jesse and Neff, Michael},
title = {Communication Behavior in Embodied Virtual Reality},
year = {2018},
isbn = {9781450356206},
publisher = {ACM},
address = {New York, NY, USA},
doi = {10.1145/3173574.3173863},
booktitle = {2018 Conference on Human Factors in Computing Systems},
pages = {1–12},
numpages = {12},
location = {Montreal QC, Canada},
series = {CHI '18}
}

@inproceedings{yun2023animationfidelityselfavatar,
  author={Yun, Haoran and Ponton, Jose Luis and Andujar, Carlos and Pelechano, Nuria},
  booktitle={2023 IEEE VR}, 
  title={Animation Fidelity in Self-Avatars: Impact on User Performance and Sense of Agency}, 
  year={2023},
  pages={286-296},
  doi={10.1109/VR55154.2023.00044}
}

@inproceedings{eubanks2020bodytrackingfidelityik,
  author={Eubanks, James Coleman and Moore, Alec G. and Fishwick, Paul A. and McMahan, Ryan P.},
  booktitle={2020 IEEE International Symposium on Mixed and Augmented Reality (ISMAR)}, 
  title={The Effects of Body Tracking Fidelity on Embodiment of an Inverse-Kinematic Avatar for Male Participants}, 
  year={2020},
  pages={54-63},
  doi={10.1109/ISMAR50242.2020.00025}
}

@INPROCEEDINGS{merz2024universalaccess,
  author={Merz, Christian and Göttfert, Christopher and Wienrich, Carolin and Latoschik, Marc Erich},
  booktitle={2024 IEEE VR}, 
  title={Universal Access for Social {XR} Across Devices: The Impact of Immersion on the Experience in Asymmetric Virtual Collaboration}, 
  year={2024},
  pages={859-869},
  doi={10.1109/VR58804.2024.00105}
}

@inproceedings{born2019exergamesimmersionembodiment,
  author={Born, Felix and Abramowski, Sophie and Masuch, Maic},
  booktitle={2019 11th VS-Games}, 
  title={Exergaming in {VR: T}he Impact of Immersive Embodiment on Motivation, Performance, and Perceived Exertion}, 
  year={2019},
  volume={},
  number={},
  pages={1-8},
  doi={10.1109/VS-Games.2019.8864579}
}

@article{goncalves2022impactbodytrackingsenseofembodiment,
author = {Gon\c{c}alves, Guilherme and Melo, Miguel and Barbosa, Lu\'{\i}s and Vasconcelos-Raposo, Jos\'{e} and Bessa, Maximino},
title = {Evaluation of the impact of different levels of self-representation and body tracking on the sense of presence and embodiment in immersive {VR}},
year = {2022},
publisher = {Springer-Verlag},
address = {Berlin, Heidelberg},
volume = {26},
number = {1},
issn = {1359-4338},
url = {https://doi.org/10.1007/s10055-021-00530-5},
doi = {10.1007/s10055-021-00530-5},
journal = {Virtual Reality},
pages = {1–14},
numpages = {14},
}

@article{hart2006rawtlx,
author = {Sandra G. Hart},
title ={{Nasa-Task Load Index (NASA-TLX);} 20 Years Later},
journal = {Proceedings of the Human Factors and Ergonomics Society Annual Meeting},
volume = {50},
number = {9},
pages = {904-908},
year = {2006},
doi = {10.1177/154193120605000909},
}

@article{kim2018vrsq,
title = {Virtual reality sickness questionnaire {(VRSQ): M}otion sickness measurement index in a virtual reality environment},
journal = {Applied Ergonomics},
volume = {69},
pages = {66-73},
year = {2018},
issn = {0003-6870},
doi = {https://doi.org/10.1016/j.apergo.2017.12.016},
author = {Hyun K. Kim and Jaehyun Park and Yeongcheol Choi and Mungyeong Choe},
}

@article{fiedler2024selfcues,
  title        = {From Avatars to Agents: Self-Related Cues through Embodiment and Personalization Affect Body Perception in Virtual Reality},
  author       = {Fiedler, Marie Luisa and Wolf, Erik and Döllinger, Nina and Mal, David and Botsch, Mario and Latoschik, Marc Erich and Wienrich, Carolin},
  journal      = {IEEE Transactions on Visualization and Computer Graphics},
  year         = {2024},
  pages        = {1-11},
  url          = {https://downloads.hci.informatik.uni-wuerzburg.de/2024-ismar-tvcg-self-identification-body-weight-perception-preprint-reduced.pdf},
  doi          = {10.1109/TVCG.2024.3456211}
}

@inproceedings{ganal2020trackingsystemcomparison,
  author = {Ganal, Elisabeth and Bartl, Andrea and Westermeier, Franziska and Roth, Daniel and Latoschik, Marc Erich},
  editor = {Hansen, C. and Nürnberger, A. and Preim, B.},
  journal = {Mensch und Computer 2020},
  publisher = {Gesellschaft für Informatik e.V.},
  title = {Developing a Study Design on the Effects of Different Motion Tracking Approaches on the User Embodiment in Virtual Reality},
  year = 2020
}

@ARTICLE{wang2023framerateux,
  author={Wang, Jialin and Shi, Rongkai and Zheng, Wenxuan and Xie, Weijie and Kao, Dominic and Liang, Hai-Ning},
  journal={IEEE Transactions on Visualization and Computer Graphics}, 
  title={Effect of Frame Rate on User Experience, Performance, and Simulator Sickness in Virtual Reality}, 
  year={2023},
  volume={29},
  number={5},
  pages={2478-2488},
  doi={10.1109/TVCG.2023.3247057}}
